\documentclass[fleqn,usenatbib]{mnras}

\usepackage{newtxtext}
\usepackage{mathptmx}
\usepackage[T1]{fontenc}

\DeclareRobustCommand{\VAN}[3]{#2}
\let\VANthebibliography\thebibliography
\def\thebibliography{\DeclareRobustCommand{\VAN}[3]{##3}\VANthebibliography}

\usepackage{graphicx}	% Including figure files
\usepackage{amsmath}	% Advanced maths commands
\usepackage{{booktabs}}
\usepackage{multirow}
\usepackage{makecell}
\usepackage{txfonts}
\usepackage{physics}
\usepackage{xcolor} % needed for colored text
\usepackage{orcidlink}

\newcommand{\xHI}{$x_{\rm HI}$}
\newcommand{\barxHI}{$\bar{x}_{\rm HI}$}

\title[Inference on 21-cm higher-order statistics]{Implicit inference of the reionization history with higher-order statistics of the 21-cm signal}

\author[N. Cerardi et al.]{Nicolas Cerardi\orcidlink{0009-0004-1864-512X},$^{1}$\thanks{E-mail: nicolas.cerardi@epfl.ch}
Sambit K. Giri\orcidlink{0000-0002-2560-536X},$^{2}$
Michele Bianco\orcidlink{0000-0002-6766-0017},$^{3}$
Davide Piras\orcidlink{0000-0002-9836-2661},$^{4}$
Emmanuel de Salis\orcidlink{0000-0002-1221-6664},$^{5}$\newauthor
Massimo De Santis\orcidlink{0009-0001-5834-3976},$^{5}$
Merve Selcuk-Simsek\orcidlink{0000-0001-6916-5899},$^{6}$
Philipp Denzel\orcidlink{0000-0003-0126-0659},$^{7}$
Kelley M. Hess\orcidlink{0000-0001-9662-9089},$^{8}$\newauthor
M. Carmen Toribio\orcidlink{0000-0001-8063-2881},$^{8}$
Franz Kirsten\orcidlink{0000-0001-6664-8668},$^{8}$
Hatem Ghorbel\orcidlink{0000-0001-5501-9807}$^{5}$
\\
$^{1}$Institute of Physics, Laboratory of Astrophysics, École Polytechnique Fédérale de Lausanne (EPFL), Switzerland\\
$^{2}$Department of Astronomy and Oskar Klein Centre, AlbaNova, Stockholm University, SE-10691 Stockholm, Sweden\\
$^{3}$Institute for Particle Physics \& Astrophysics (ETHZ), Wolfgang-Pauli-Str 27, 8093 Zurich, Switzerland \\
$^{4}$Département de Physique Théorique, Université de Genève, 24 quai Ernest Ansermet, 1211 Genève 4, Switzerland \\
$^{5}$Haute Ecole Arc Ing\'enierie, University of Applied Sciences \& Arts Western Switzerland (HES-SO), Saint-Imier, Switzerland \\
$^{6}$Institute for Data Science, University of Applied Sciences \& Arts Northwestern Switzerland (FHNW), Bahnhofstrasse 6, Windisch, 5210, Switzerland \\
$^{7}$Centre for Artificial Intelligence, Zurich University of Applied Sciences (ZHAW), Technikumstrasse 71, 8400 Winterthur, Switzerland \\
$^{8}$ Department of Space, Earth and Environment, Chalmers University of Technology, Onsala Space Observatory, SE-43992 Onsala, Sweden
}

\date{Accepted 2026 April 18. Received 2026 March 20; in original form 2025 November 21}

\pubyear{\the\year{}}

\begin{document}
\label{firstpage}
\pagerange{\pageref{firstpage}--\pageref{lastpage}}
\maketitle

\begin{abstract}
The Epoch of Reionization (EoR), when the first luminous sources ionised the intergalactic medium, represents a new frontier in cosmology. The Square Kilometre Array Observatory (SKAO) will offer unprecedented insights into this era through observations of the redshifted 21-cm signal, enabling constraints on the Universe's reionization history. We investigate the information content of the average neutral hydrogen fraction ($\bar{x}_{\rm HI}$) in several Gaussian (spherical and cylindrical power spectra) and non-Gaussian (Betti numbers and bispectrum) summary statistics of the 21-cm signal. 
Mock 21-cm observations are generated using the AA* configuration of SKAO's low-frequency telescope, incorporating noise levels for 100 and 1000 hours.
We employ a state-of-the-art implicit inference framework to learn posterior distributions of $\bar{x}_{\rm HI}$ in redshift bins centred at $z=8.0,7.2$ and 6.5, for each statistic and noise scenario, validating the posteriors through calibration tests. Using the figure of merit to assess constraining power, we find that Betti numbers alone are on average more informative than the power spectra, while the bispectrum provides limited constraints. However, combining higher-order statistics with the cylindrical power spectrum improves the mean figure of merit by $\sim$0.25 dex ($\sim$33\% reduction in $\sigma(\bar{x}_{\rm HI})$). The relative contribution of each statistic varies with the stage of reionization.  With SKAO observations approaching, our results show that combining power spectra with higher-order statistics can significantly increase the information retrieved from the EoR, maximising the scientific return of future 21-cm observations.

\end{abstract}


%%%%%%%%%%%%%%%%%%%%%%%%%%%%%%%%%%%%%%%%%%%%%%%%%%

%%%%%%%%%%%%%%%%% BODY OF PAPER %%%%%%%%%%%%%%%%%%

\section{Introduction}
The Epoch of Reionization (EoR) marks one of the most significant phase transitions in cosmic history. During this period, the first generation of stars and galaxies formed, and their ultraviolet radiation progressively ionised the neutral hydrogen (HI) that filled the intergalactic medium (IGM). This process, which concluded approximately one billion years after the Big Bang, brought an end to the cosmic ``dark ages'' and fundamentally shaped the universe we observe today. A primary goal of modern cosmology is to chart the timeline of this process, typically parametrised by the evolution of the global neutral hydrogen fraction \barxHI{}. Observations of the Lyman-$\alpha$ forest in high-redshift quasar spectra have suggested that reionization ended around $z \approx 5.3-5.5$ \citep[e.g.,][]{fan2006observational,becker2015evidence,kulkarni2019large,bosman2022hydrogen}, and several complementing constraints from the CMB optical depth \citep[][]{planckcollaboration2020results} and galaxy observations \citep[e.g.,][]{mason2018universe,mason2019inferences,hoag2019constraining,bolan2022inferring,jones2025jades} indicate that it was halfway complete around $z \approx 7$.

While powerful, these existing observational probes offer an indirect view of the IGM during reionization. Quasar absorption spectra, galaxy damping wings, and Lyman-$\alpha$ emitter statistics rely on interpreting the imprint of the IGM on light originating from or passing near luminous sources \citep[e.g.,][]{mcgreer2015model,davies2018quantitative,greig2022igm,curtis2023spectroscopic,umeda2024jwst}. Inferring the global properties of the IGM from these sightlines requires careful modelling of both the sources and the complex radiative transfer of Lyman-$\alpha$ photons through an inhomogeneous medium \citep[e.g.,][]{huberty2025pitfalls,georgiev2025forest}. Furthermore, inferring global quantities from a handful of sightlines can be prone to cosmic variance. In contrast, the 21-cm signal, a spectral line from the hyperfine spin-flip transition of neutral hydrogen, offers a more direct window into the IGM during the EoR \citep[see e.g.,][for a review]{pritchard201221}. Originating from the neutral hydrogen distributed throughout the cosmic volume, mapping the 21-cm signal across different redshifts can provide a three-dimensional tomography of the IGM, revealing the evolving morphology and topology of the ionised bubbles \citep[e.g.,][]{friedrich2011topology,giri2018bubble,giri2021measuring}. This, in turn, allows us to constrain not only the underlying astrophysical processes but also the reionization history directly \citep[e.g.,][]{giri2018optimal,bianco2021deep,bianco2024deep}.
However, detecting this faint signal is an extreme observational challenge. It is buried beneath astrophysical foregrounds that are orders of magnitude brighter, requiring exceptionally precise calibration \citep[e.g.,][]{barry2016calibration,mevius2022numerical,gan2023assessing}. Current-generation radio interferometers, such as the Low Frequency Array \citep[LOFAR;][]{mertens2020improved,mertens2025deeper}, the Murchison Widefield Array \citep[MWA;][]{trott2020deep,nunhokee2025limits}, and the Hydrogen Epoch of Reionization Array \citep[HERA;][]{hera2022first,hera2023improved}, have made significant progress, placing the first upper limits on the 21-cm power spectrum. These data have already been used to rule out some extreme scenarios, particularly those involving minimal heating of the IGM as reionization proceeds (`cold reionization') which predict a large 21-cm signal amplitude \citep[e.g.,][]{greig2021interpreting,greig2021exploring,ghara2020constraining,ghara2021constraining,ghara2025constraints,hera2023improved,nunhokee2025limits}.

The forthcoming Square Kilometre Array’s Low-frequency component \citep[SKA-Low;][]{koopmans2015cosmic} is poised to revolutionise the field. With its unprecedented sensitivity, SKA-Low is expected to move beyond upper limits to achieve a robust detection and detailed characterisation of the 21-cm signal across cosmic time. A primary scientific objective is to leverage these data to directly reconstruct the reionization history. The cylindrically averaged power spectrum (PS2D) serves as the primary summary statistic for initial analyses, offering robustness against foreground contamination and being the expected first statistical detection \citep[e.g.,][]{liu2011method,pober2014next}. Indeed, forecast studies have demonstrated the potential of the PS2D to constrain the reionization history \citep[e.g.,][]{greig2024inferring,cooper2025simulation,pietschke2025direct}. The importance of this statistic and objective was also highlighted in the recent SKA Science Data Challenge 3b (SDC3b\footnote{\url{https://sdc3.skao.int/}}), where international teams competed to infer the reionization history from blinded mock PS2D data. Among the participants, the SEarCH (\textbf{S}vea's \textbf{E}ndevour in \textbf{A}I and \textbf{R}eionization \textbf{C}osmology with \textbf{H}elvetia) team provided two methods \citep[][and this work]{de2025exploring}, both securing podium positions in the SDC3b rankings.

The 21-cm signal, however, is highly non-Gaussian, reflecting the complex and evolving morphology of ionised regions \citep[e.g.,][]{iliev2006simulating,majumdar2018quantifying,giri2018bubble,giri2019position}. The growth of ionised bubbles, a key tracer of the reionization history, introduces significant non-Gaussian features in the signal's morphology and topology. The PS2D, as a two-point statistic, only captures the complete information for a Gaussian field and is therefore insufficient, potentially missing a wealth of information. This strongly motivates the exploration of non-Gaussian information through higher-order summary statistics, such as the bispectrum (a three-point statistic) and topological descriptors such as Betti numbers \citep[e.g.,][]{elbers2019persistent, giri2021measuring, kapahtia2021prospects}, which are explicitly sensitive to the complex morphology of the ionised regions. In this study, we explore the capabilities of the non-Gaussian information on constraining the reionization history.

Building an inference framework capable of optimally extracting information from the 21-cm signal data faces significant hurdles. Firstly, modelling the EoR and the resulting 21-cm signal is computationally demanding. Even fast semi-numerical codes such as \textsc{21cmFAST} \citep[][]{mesinger201121cmfast}, \textsc{GRIZZLY} \citep[][]{ghara2018prediction} and \textsc{BEoRN} \citep[][]{schaeffer2023beorn} require several CPU hours for a single simulation spanning the relevant cosmic time ($5 \lesssim z \lesssim 12$). Traditional Markov chain Monte Carlo (MCMC) frameworks, which require evaluating millions of models to explore the high-dimensional astrophysical parameter space adequately, become computationally prohibitive \citep[e.g.,][]{greig201521cmmc}. Secondly, the likelihood function $p(\mathbf{t}|x_{\rm HI}, \theta)$, which connects $\mathbf{t}$, the summary statistic of the observed data, to the reionization state $x_{\rm HI}$ and underlying parameters $\theta$, is often intractable, especially for higher-order statistics or when complex instrumental effects are considered \citep[][]{prelogovic2023exploring}. Moreover, previous work has demonstrated how combining multiple complementary summary statistics can maximise the astrophysical inference from 21-cm data \citep{Prelogovic2024howinform}. Employing fast \textit{emulators} trained on simulation suites have been proposed to accelerate the forward modelling of specific summary statistics within MCMC \citep[e.g.,][]{kern2017emulating,Schmit2018emulation,ghara2020constraining,breitman202421cmemu,Meriot2024loreli,Choudhury2024gprSCRIPT}. While this strategy can address the computational cost after an initial training investment, emulators are typically specific to the statistic they were trained on and require retraining for new observables; crucially, they do not inherently solve the problem of an intractable or unknown likelihood function. 

Implicit inference, also known as simulation-based inference (SBI), offer a compelling alternative that circumvents the need for an explicit analytical form for the likelihood $p(\mathbf{t}|\theta)$ \citep[see e.g.,][for a review]{Cranmer2020fontierSBI}. These techniques leverage the fact that complex EoR models can generate simulated data instances which carry an implicit likelihood by sampling the joint distribution ($\theta$, $\mathbf{t}$). By training neural networks on a sufficiently large set of pre-computed simulations, SBI algorithms learn approximations of e.g. the posterior $p(\theta|\mathbf{t})$, the likelihood $p(\mathbf{t}|\theta)$, or the likelihood ratio $p(\mathbf{t}|\theta) / p(\mathbf{t}|\theta_{\text{ref}})$, where $\theta_{\text{ref}}$ is a chosen fixed reference parameter set. These techniques enable the use of complex simulation-based models within inference tasks, ultimately promising to fully exploit the information contained in observations and tackling systematics. As such, implicit inference has been tested for a wide variety of tasks in cosmology and astrophysics \citep[e.g.,][]{Vasist_2023, Lanzieri_2025, CerardiCluster2025, Saoulis_2025}. Implicit inference provides particular advantages for 21-cm cosmology. First, it naturally handles the intractable likelihoods associated with non-Gaussian summary statistics or complex instrumental effects. Furthermore, the methodology efficiently utilises computationally expensive simulations, as the generation of the training set can be parallelised and generally requires significantly fewer simulation runs than a typical MCMC chain needs to converge \citep[e.g.,][]{Schmit2018emulation,Semelin2025combining}. Consequently, SBI methods are increasingly being applied to 21-cm signal analysis. Studies have employed various summary statistics, including the power spectrum, PDF moments, and wavelet scattering transforms, within implicit inference frameworks to forecast constraints on astrophysical parameters \citep[e.g.,][]{Zhao2022delfi,Zhao2024wavelet,prelogovic2023exploring,greig2024inferring,Meriot2025bayesian}. Recently, \citet{Semelin2025combining} have demonstrated the capability of SBI to combine information from multiple, potentially correlated, summary statistics to derive tighter constraints.

In this paper, we study the non-Gaussian information in the 21-cm signal to improve the constraints on reionization history. This paper is structured as follows. In Sec.~\ref{sec:data_model}, we introduce the adopted forward model, including semi-analytical simulations, instrumental effects, and the chosen summary statistics for this study. In Sec.~\ref{sec:inference}, we detail our implicit inference framework to derive posterior estimators from single and combined summary statistics. In Sec.~\ref{sec:results}, we present our results, showing calibration metrics and comparing summary statistics performance. Lastly, in Sec.~\ref{sec:conclusion} we discuss our results and conclusions from this work. Throughout this article, we assume a flat $\Lambda$CDM cosmology, with the \cite{planckcollaboration2020results} parameters: $H_0=67.66$ km s$^{-1}$ Mpc$^{-1}$, $\Omega_{\rm m} = 0.30964$, $\sigma_8 = 0.8102$ and $n_s = 0.9665$.

\section{Data model}\label{sec:data_model}

\subsection{Simulations}
Our inference framework is trained on a large suite of mock 21-cm signal observations. In this section, we describe the numerical simulations used to generate this training set. We first detail the simulation code and the astrophysical parameterisation adopted (Sec.~\ref{sec:sim_method}), and then define the three specific reference models used to test our framework (Sec.~\ref{sec:sim_cases}).

\subsubsection{Method}\label{sec:sim_method}
To generate the training data for our SBI framework, we employ the widely-used semi-numerical simulation code \textsc{21cmFAST} \citep[][]{mesinger201121cmfast,murray202021cmfast}. \textsc{21cmFAST} efficiently models the large-scale structure formation using second-order Lagrangian Perturbation Theory \citep[2LPT; e.g.,][]{scoccimarro1998transients} and simulates the process of reionization using an excursion-set formalism based on \citet{furlanetto2004growth}. This approach allows for the rapid generation of large cosmological volumes necessary for capturing the statistical properties of the 21-cm signal. The code self-consistently calculates the evolution of density, velocity, ionization, and temperature fields, enabling the computation of the 21-cm brightness temperature contrast.

The observable 21-cm signal is quantified by the differential brightness temperature $\delta T_b$, given as \citep[e.g.,][]{furlanetto2006cosmology}:
\begin{eqnarray}
    \delta T_{\rm b} (\mathbf{x}, z) \approx 27 x_{\rm HI}(\mathbf{x}, z) [1 + \delta_{\rm b}(\mathbf{x}, z)] \left( \frac{\Omega_{\rm b} h^2}{0.023} \right) \left( \frac{0.15}{\Omega_{\rm m} h^2} \frac{1+z}{10} \right)^{1/2} \nonumber \\
    \left( 1 - \frac{T_\gamma(z)}{T_S(\mathbf{x}, z)} \right) \left( \frac{H(z)}{dv_r/dr + H(z)} \right) \rm{mK}
    \label{eq:dTb}
\end{eqnarray}
where $x_{\rm HI}(\mathbf{x}, z)$ and $\delta_{\rm b}(\mathbf{x}, z)$ are the neutral hydrogen fraction and baryon overdensity, respectively, at position $\mathbf{x}$ and redshift $z$. $T_S(\mathbf{x}, z)$ is the spin temperature, determined by the coupling of the hyperfine levels to the CMB, the kinetic gas temperature (via collisions), and the Lyman-$\alpha$ background (via Wouthuysen-Field effect). $T_\gamma(z)$ is the CMB temperature at redshift $z$. $\Omega_{\rm b}$ and $\Omega_{\rm m}$ are the baryon and total matter density contrast parameters, respectively. $dv_r/dr$ is the gradient of the peculiar velocity along the line-of-sight, accounting for redshift-space distortions.

For this work, we use the full spin temperature calculation within \textsc{21cmFAST}, which accounts for collisional coupling, Wouthuysen-Field coupling from Lyman-$\alpha$ photons, and heating from X-ray photons. This provides a more realistic description than assuming the high spin temperature limit ($T_S \gg T_\gamma$), especially at earlier times or in potentially colder regions of the IGM. Although our primary focus is on the later stages of reionization ($z \lesssim 9$), including the full spin temperature evolution ensures our framework can handle variations in thermal state and is robust even if some regions remain relatively cold, allowing us to test the framework's ability to constrain the reionization history across diverse scenarios.

\begin{table*}
\caption{The astrophysical parameter values of the three reference scenarios from our testing set, along with the corresponding sky-averaged neutral fractions within $\sim$15 MHz frequency band at the three redshifts considered in this study.}             % title of Table
\label{table:params_3models}      % is used to refer this table in the text
\centering                          % used for centering table
\begin{tabular}{c c c c c c c c c c}        % centered columns (4 columns)
\hline\hline                 % inserts double horizontal lines
Model & log$_{10}f_{\star,10}$ & $\alpha_\star$ & log$_{10}f_{esc,10}$ & $\alpha_{esc}$ & log$_{10}M_{turn}$ & $t_\star$ & $\bar{x}_{\rm HI}(z=8.0)$ & $\bar{x}_{\rm HI}(z=7.2)$ & $\bar{x}_{\rm HI}(z=6.5)$\\    % table heading 
\hline                        % inserts single horizontal line
   % Early    & -1.10 & -0.29 & -0.73 & -0.58 & 9.71 & 0.36 & 0.41 & 0.13 & 0.00 \\  %ID 15234 
   Early    & -1.05 & -0.01 & -1.01 & -0.91 & 9.60 & 0.60 & 0.46 & 0.21 & 0.02 \\  %ID 15442 
   Fiducial & -0.38 & -0.21 & -2.11 &  0.31 & 8.63 & 0.72 & 0.59 & 0.43 & 0.20 \\ %ID 15089 
   Late     & -0.43 &  0.21 & -2.25 &  0.05 & 8.22 & 0.19 & 0.70 & 0.56 & 0.39 \\ %ID 15232 
\hline                                   %inserts single line
\end{tabular}
\end{table*}

\textsc{21cmFAST} models the properties of galaxies, which drive reionization and IGM heating, using a set of astrophysical parameters linked to the mass of dark matter halos ($M_h$), following the parametrisation detailed in \citet{park2019inferring}. The star formation efficiency (SFE), representing the fraction of galactic gas converted into stars, is parameterised by a normalisation $f_{*,10}$ at a halo mass of $10^{10} M_{\odot}$ and a power-law index $\alpha_{*}$ describing its scaling with halo mass ($f_{*} \propto M_h^{\alpha_{*}}$). Similarly, the escape fraction of ionising UV photons, $f_{\rm esc}$, which dictates how many ionising photons reach the IGM, is modelled with a normalisation $f_{\rm esc,10}$ and a power-law index $\alpha_{\rm esc}$. The ability of low-mass halos to form stars is regulated by $M_{\rm turn}$ below which star formation is exponentially suppressed due to feedback or cooling limitations. The timescale over which galaxies form their stars is parameterised by $t_{*}$, expressed as a fraction of the Hubble time. IGM heating is primarily driven by X-rays, modelled by the soft-band (0.5-2 keV) X-ray luminosity per unit star formation rate, $L_{X<2{\rm keV}}/{\rm SFR}$, which sets the overall X-ray efficiency. The shape of the X-ray spectrum emergent from galaxies is controlled by $E_0$, the minimum energy (in keV) required for X-ray photons to escape absorption within the host galaxy, and $\alpha_X$, the power-law spectral index ($L_X \propto E^{-\alpha_X}$). In principle, varying all nine parameters ($f_{*,10}, \alpha_{*}, f_{\rm esc,10}, \alpha_{\rm esc}, M_{\rm turn}, t_{*}, L_{X<2{\rm keV}}/{\rm SFR}, E_0, \alpha_X$) allows for a wide exploration of signal characteristics. We describe our prior sampling strategy in the following sub-section.

\subsubsection{Reference cases}\label{sec:sim_cases}

\begin{figure}
    \centering
    \includegraphics[width=\linewidth, trim=0.5cm 0.5cm 0 0, clip]{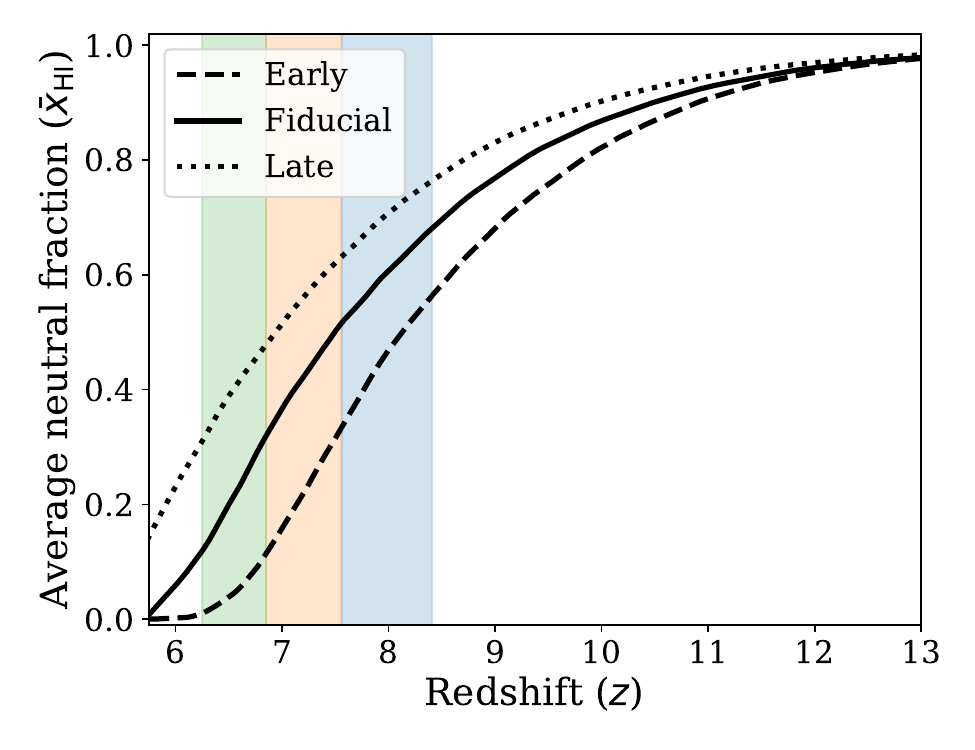}
    \caption{Redshift evolution of sky-averaged neutral fraction \barxHI of three scenarios (early, fiducial and late reionization models). The three coloured bands show the frequency bins considered in this study.
    }
    \label{fig:eor_hist}
\end{figure}

To test the performance of our inference framework, we define three reference scenarios corresponding to an `Early', `Fiducial', and `Late' reionization history. The astrophysical parameters defining these three models are taken from our test set and are detailed in Table~\ref{table:params_3models}. The resulting evolution of the sky-averaged neutral fraction $\bar{x}_{\rm HI}$ for each case is shown in Fig.~\ref{fig:eor_hist}. The `Fiducial' model's reionization history begins at $z \approx 13$ and ends by $z \approx 5.5$, which is consistent with current constraints on reionization \citep[see e.g.,][for discussion about these constraints]{giri202421,qin2025percent}. The other two models represent more extreme scenarios. While all models begin at a similar epoch, reionization in the `Early' model proceeds very rapidly and is nearly complete by $z\approx 6.5$ with a neutral fraction of $\bar{x}_{\rm HI} \approx 0.02$. Conversely, the 'Late' model shows a more gradual trend, remaining significantly neutral with an average neutral fraction $\bar{x}_{\rm HI} \gtrsim 0.1$ at $z \approx 6$. The specific values of the neutral fraction for each model at our three redshifts of interest ($z=8.0, 7.2, \text{and } 6.5$) are listed in Table~\ref{table:params_3models}.

To properly model the signal for such evolutionary histories, we generate \textit{lightcone} data volumes from our \textsc{21cmFAST} models. A \textit{lightcone} data represents the 21-cm signal as it evolves along the line-of-sight, since different observed frequencies, $\nu_\mathrm{obs}$, correspond to different redshifts (and thus different cosmic times). For our analysis, we focus on measurements within three distinct redshift bins centred at $z=8.0,\,7.2$, and 6.5 ($\nu_\mathrm{obs}\simeq158,\,173$ and $189$ MHz), each with a bandwidth of $\sim$15 MHz. These bins are highlighted by the coloured bands in Fig.~\ref{fig:eor_hist}. While the signal evolution within this bandwidth introduces the lightcone effect \citep[e.g.,][]{datta2012lightcone,giri2018bubble}, our explicit construction of lightcones for all models in the training set ensures that this effect is consistently included and does not introduce bias into our inference framework.

\subsection{Sampling prior}\label{sec:sampling_prior}

\begin{figure}
    \centering
    \includegraphics[width=\linewidth, trim=2cm 0 2cm 0, clip]{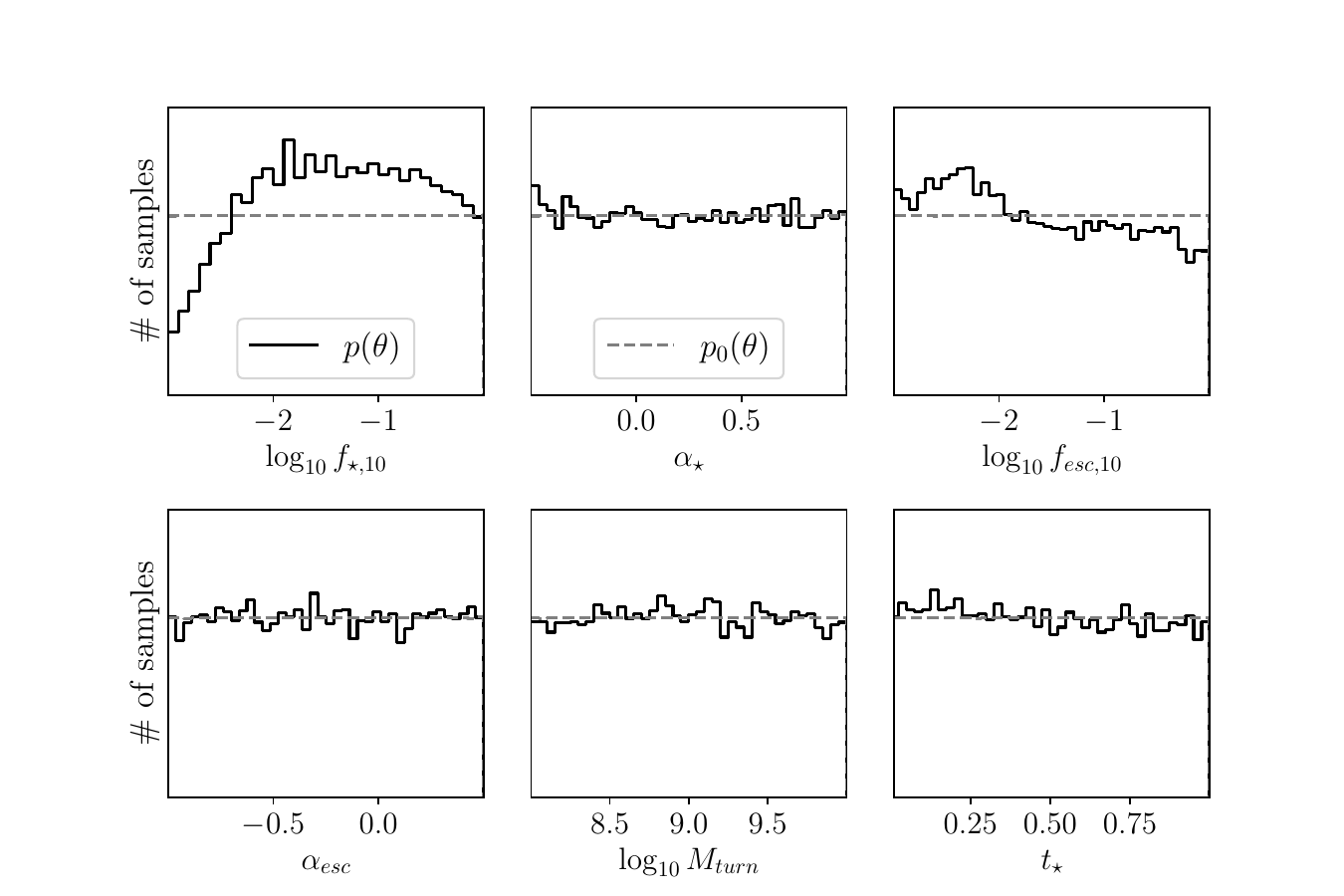}
    \includegraphics[width=\linewidth, trim=2cm 0 2cm 0.5cm, clip]{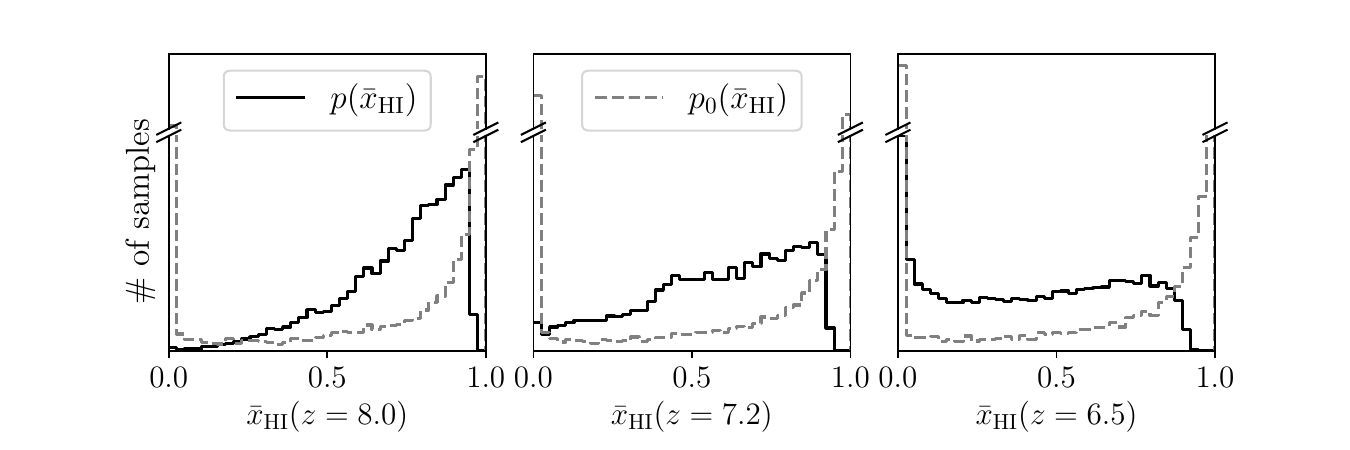}
    \caption{Marginal distributions on the simulation parameters $\theta$(top and middle row) and on the resulting \barxHI{} for three different redshifts (bottom row). \textit{y}-axes show the number of sample in the dataset. A uniform sampling $p_0(\theta)$ of the astrophysical parameters (dashed histograms) give a highly bimodal distribution $p_0(\bar{x}_{\rm HI})$ of \barxHI{} in the three frequency bins. With our custom sampling strategy $p(\theta)$ (see Sec.~\ref{sec:sampling_prior}), we get a relatively more balanced distribution $p(\bar{x}_{\rm HI})$ of these bins (solid histograms).
    }
    \label{fig:samplingpriors}
\end{figure}

For this study, we vary the six key astrophysical parameters $\theta$ controlling the galaxy UV properties and star formation activity: $f_{*,10}, \alpha_{*}, f_{\rm esc,10}, \alpha_{\rm esc}, M_{\rm turn}, t_{*}$. The remaining three parameters describing the X-ray properties of galaxies ($L_{X<2{\rm keV}}/{\rm SFR}$, $E_0$, $\alpha_X$) were kept fixed to the fiducial values recommended in \citet{park2019inferring}. It is important to note that the heating of the IGM is still modulated through $f_{*,10}$, $\alpha_{*}$ and $M_{\rm turn}$. For example, heating could be delayed if star formation is very low (low $f_{*,10}$) and only occurs in massive haloes (high $M_{\rm turn}$). We initially sampled the six varied parameters from broad, uniform prior distributions: $\log_{10}(f_{*,10}) \in [-3, 0]$, $\alpha_{*} \in [-0.5, 1]$, $\log_{10}(f_{\rm esc,10}) \in [-3, 0]$, $\alpha_{\rm esc} \in [-1, 0.5]$, $\log_{10}(M_{\rm turn}/M_\odot) \in [8, 10]$, and $t_{*} \in (0, 1]$. The initial sampling distributions $p_0(\theta)$ for these astrophysical parameters are shown as dashed grey lines in the top two rows of Fig. \ref{fig:samplingpriors}. However, this uniform sampling in the astrophysical parameter space leads to a highly non-uniform, bimodal distribution of the resulting reionization histories, $\bar{x}_{\rm HI}(z)$, as illustrated by the dashed grey lines in the bottom row of Fig. \ref{fig:samplingpriors} for $z=8.0, 7.2,$ and $6.5$. This bimodality, with peaks corresponding to very early/rapid or very late/slow reionization scenarios, presents challenges for training neural networks in SBI frameworks, potentially leading to inefficient learning or biased posteriors.

To address this, we implemented a custom sampling strategy designed to yield a more balanced distribution of reionization histories. We leveraged the publicly available \textsc{21cmEMU} emulator \citep[][]{breitman202421cmemu}, which provides a rapid mapping from the \textsc{21cmFAST} astrophysical parameters to the evolution of the sky-averaged neutral fraction, $\bar{x}_{\rm HI}$. We drew parameter sets $\theta$ from the initial uniform priors $p_0(\theta)$ and used \textsc{21cmEMU} to predict the corresponding $\bar{x}_{\rm HI}$. We then applied rejection sampling, discarding models where reionization either completed too early (defined as $\bar{x}_{\rm HI}(z=8.4) \leq 0.01$) or had not significantly progressed by the end of our redshift range of interest (defined as $\bar{x}_{\rm HI}(z=6.25) \geq 0.95$). Every few hundred iterations, we stop the parameter sampling and modify the upper and lower limits to target specific reionization scenarios, aiming to obtain a more balanced distribution of $\bar{x}_{\rm HI}$ across the three redshift bins. Nevertheless, we can observe that the highest (lowest) redshift has a preference for late (early) scenarios. We continued this process until we collected $\sim$15,900 valid parameter sets. These sets were then used to run full \textsc{21cmFAST} simulations to generate our final training dataset. We perform each simulation in a $(128\times 3)^3$ grid box with $500$ Mpc along each side, and then downsize to $(128)^3$ voxels to avoid small-scale effects due to the semi-numerical framework \citep[see e.g.,][for discussions about these effects]{choudhury2018photon}. While we expect the sample variance to be small in our case for 2-point statistics ($~5\%$ at $k=0.1\ h^{-1}$Mpc with our chosen box size, as shown by \citealt{giri2023suppressing}), its effect on higher-order statistics is unclear. Hence, we vary the cosmological initial conditions for each sample to consistently include the sample variance in this dataset. As a result, the trained neural estimators are marginalizing over the initial conditions, and the sample variance is included in the predicted posteriors.

The effective sampling priors $p(\theta)$ resulting from this procedure are shown as solid dark histograms in the top two rows of Fig. \ref{fig:samplingpriors}. While no longer strictly uniform, they still cover the initial parameter ranges broadly. Crucially, the distribution of neutral fractions $p(\bar{x}_{\rm HI})$ resulting from this custom sample, shown as solid dark histograms in the bottom row of Fig. \ref{fig:samplingpriors}, is significantly more balanced and less bimodal compared to the uniform astrophysical prior case. Although not perfectly uniform in $\bar{x}_{\rm HI}$, this improved distribution provides a much better basis for training our inference framework.

\subsection{Forward-modelling instrumental effects}\label{sec:instrumental_effects}

Implicit inference methods critically rely on the fidelity of the training set in mimicking the observation. Realistic forward-modelling of observational contaminants and instrumental effects is crucial for the neural networks to learn the correct mapping between the data summaries to underlying parameters, especially when the likelihood function is intractable. In this work, we incorporate the impact of instrumental noise and finite resolution characteristic of SKA-Low observations.

We consider two mock observational setups corresponding to the AA* antenna configuration of SKA-Low, which includes 307 antenna stations. The initial science viable data will be produced with this configuration. The first setup assumes a total integration time of 100 hours, while the second assumes a deeper observation of 1000 hours. For both setups, we assume 6 hours of observation per day, centred on the field transiting zenith at a declination of $-30$ degrees. Instrumental noise is a primary contaminant and we calculate the expected noise standard deviation per visibility ($\sigma_{\rm noise}$) as following \citep[e.g.,][]{ghara2017imaging}:
\begin{eqnarray}
    \sigma_{\rm noise} = \frac{\text{SEFD}}{\sqrt{2 \Delta \nu \Delta t}}
\end{eqnarray}
where $\Delta \nu$ is the channel width and $\Delta t$ is the integration time per visibility. The SEFD is the system equivalent flux density that depends on the system temperature $T_{\rm sys}$ and the effective area $A_{\rm eff}$ of each station (SEFD = $2 k_B T_{\rm sys} / A_{\rm eff}$, where $k_B$ is the Boltzmann constant). We adopt the frequency-dependent $T_{\rm sys}$ and $A_{\rm eff}$ models relevant for SKA-Low, as defined in \citet{giri2018optimal}. Following the procedure in \citet{giri2018optimal}, we generate Gaussian random noise realisations on a grid matching the \textsc{21cmFAST} simulation resolution, consistent with this $\sigma_{\rm noise}$. The final noise rms at the map level for 100h (1000h) of observation is of $\sim160$, $230$, $320$ mK ($\sim50$, $70$, $100$ mK), respectively at $z=6.5, 7.2, 8.0$.

The interferometer does not sample all spatial frequencies uniformly; its sampling pattern in the uv-plane is determined by the station layout and Earth's rotation. Using the publicly available package \textsc{Tools21cm} \citep[][]{giri2020tools}, we simulate the daily uv-coverage tracks based on the SKA-Low AA* station positions. The simulated noise grids are then convolved with the weights derived from these uv-tracks. We assume a natural weighting scheme, which optimally weights the visibilities based on sampling density to minimise the noise variance in the final map, thereby maximising sensitivity.

Furthermore, the finite extent of the interferometer limits the angular resolution. We mimic this by filtering the simulated signal and noise fields in Fourier space, retaining only the modes corresponding to baselines shorter than the maximum baseline of the SKA-Low core ($\sim$1 km diameter). This effectively reduces the angular resolution of the data. To ensure isotropic resolution, we also match the resolution along the frequency axis by smoothing the data cubes to have approximately the same physical scale per voxel as in the angular directions after filtering.

A crucial aspect of interferometry is the absence of a zero-spacing baseline, meaning the sky-averaged signal within each frequency channel cannot be measured. Therefore, before computing any summary statistics, we subtract the spatial mean from each frequency slice of our simulated data cubes (signal plus noise) to mimic this observational characteristic.
This procedure of generating noise on a grid and then convolving with the uv-weights is computationally efficient compared to direct visibility simulation, allowing us to rapidly produce multiple noise realisations.  To capture this stochasticity in our training set, we generate approximately 25 independent noise realisations for each \textsc{21cmFAST} model lightcone. These realisations aid the implicit inference framework learn to marginalise over the noise properties during inference.

\subsection{Summary statistics}\label{sec:summary_stats}

\begin{figure*}
    \centering
    \includegraphics[width=0.82\textwidth, trim=0.5cm 0.75cm 0.5cm 0.6cm, clip]{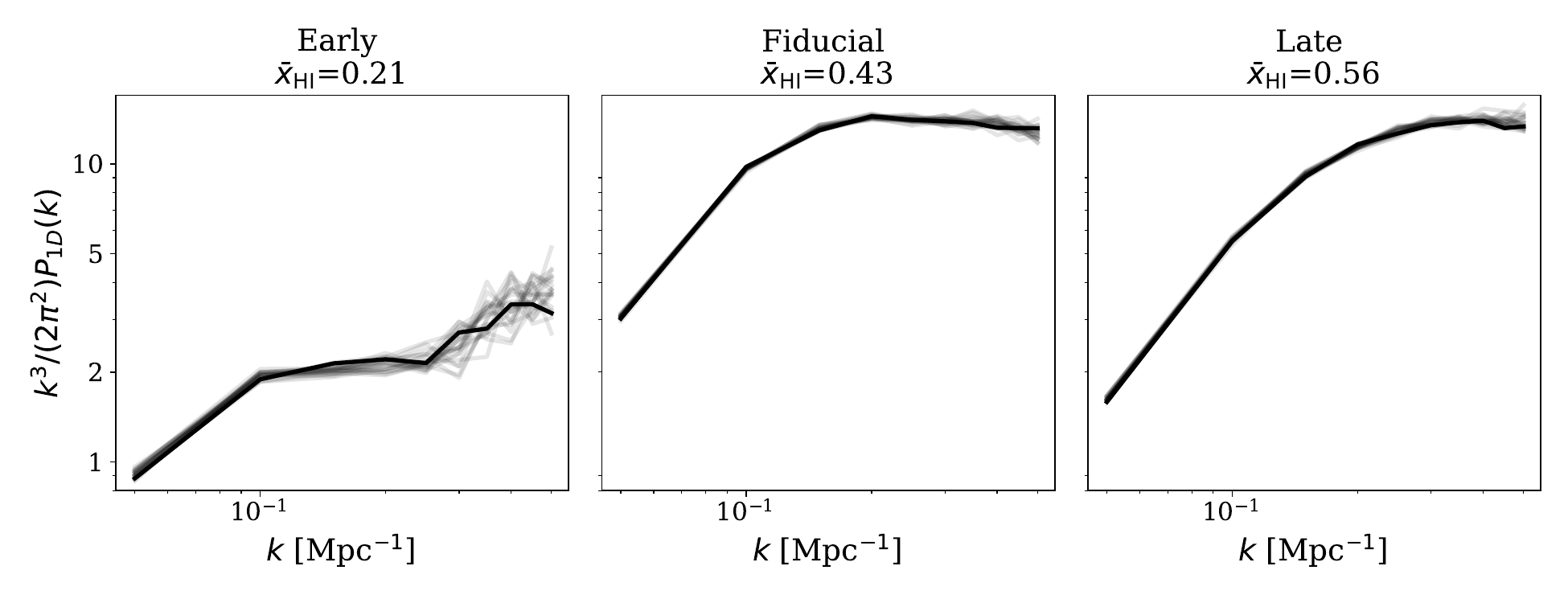}
    \includegraphics[width=0.98\textwidth, trim=-2.5cm 0.cm 0.cm 0.cm, clip]{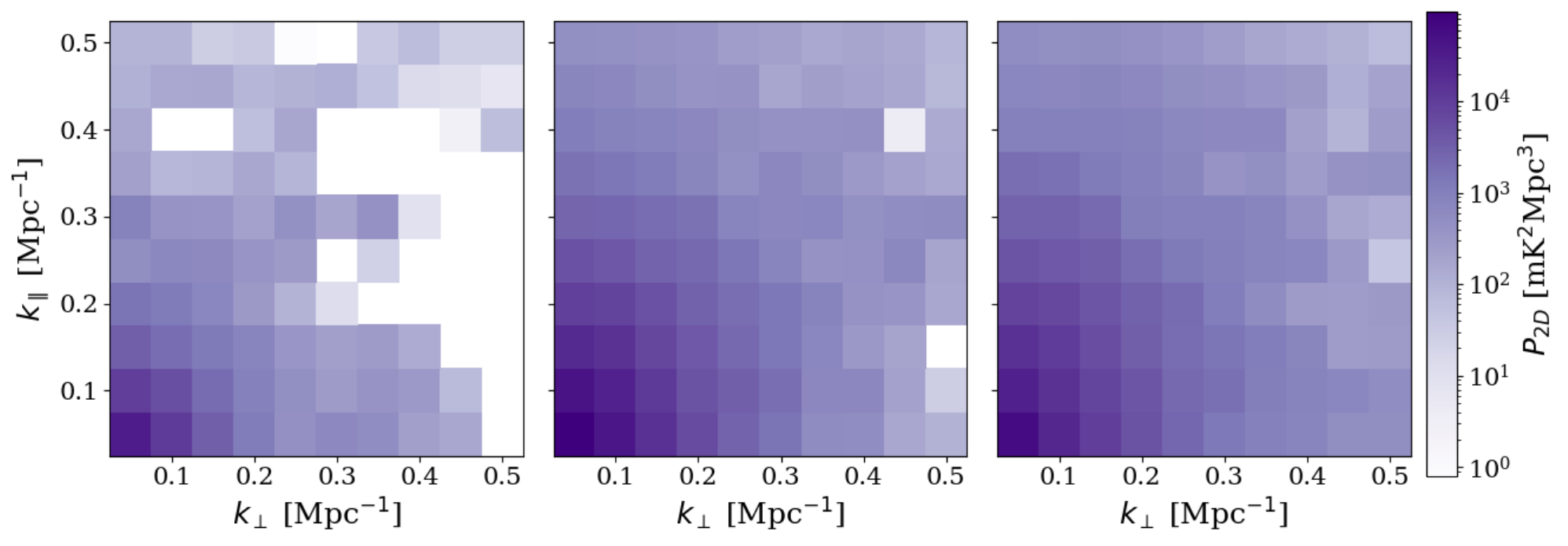}
    \caption{
    Gaussian summary statistics at $z=7.2$ for the three reference models (left to right: Early, Fiducial, Late), assuming 1000h SKA-Low noise. \textit{Top row:} The spherically averaged power spectrum, $P_{1D}(k)$. Lighter lines show multiple noise realisations, and the darker line highlights a random one. The instrumental noise bias has been removed from all curves for clarity. \textit{Bottom row:} The cylindrically averaged power spectrum, $P_{2D}(k_{\perp}, k_{\parallel})$, for the single realisation highlighted in the top row, again with the noise bias removed. White pixels (e.g., in the `Early' model) indicate noise-dominated modes where bias subtraction resulted in negative values.
    }
    \label{fig:stat_examples_gaussian}
\end{figure*}

\begin{figure*}
    \centering
    \includegraphics[width=0.85\textwidth, trim=0.0cm 10.8cm 0.0cm 0.6cm, clip]{figures/PS1D_3models.pdf}
    \includegraphics[width=0.80\textwidth, trim=0.0cm 0.0cm 0.0cm 0.2cm, clip]{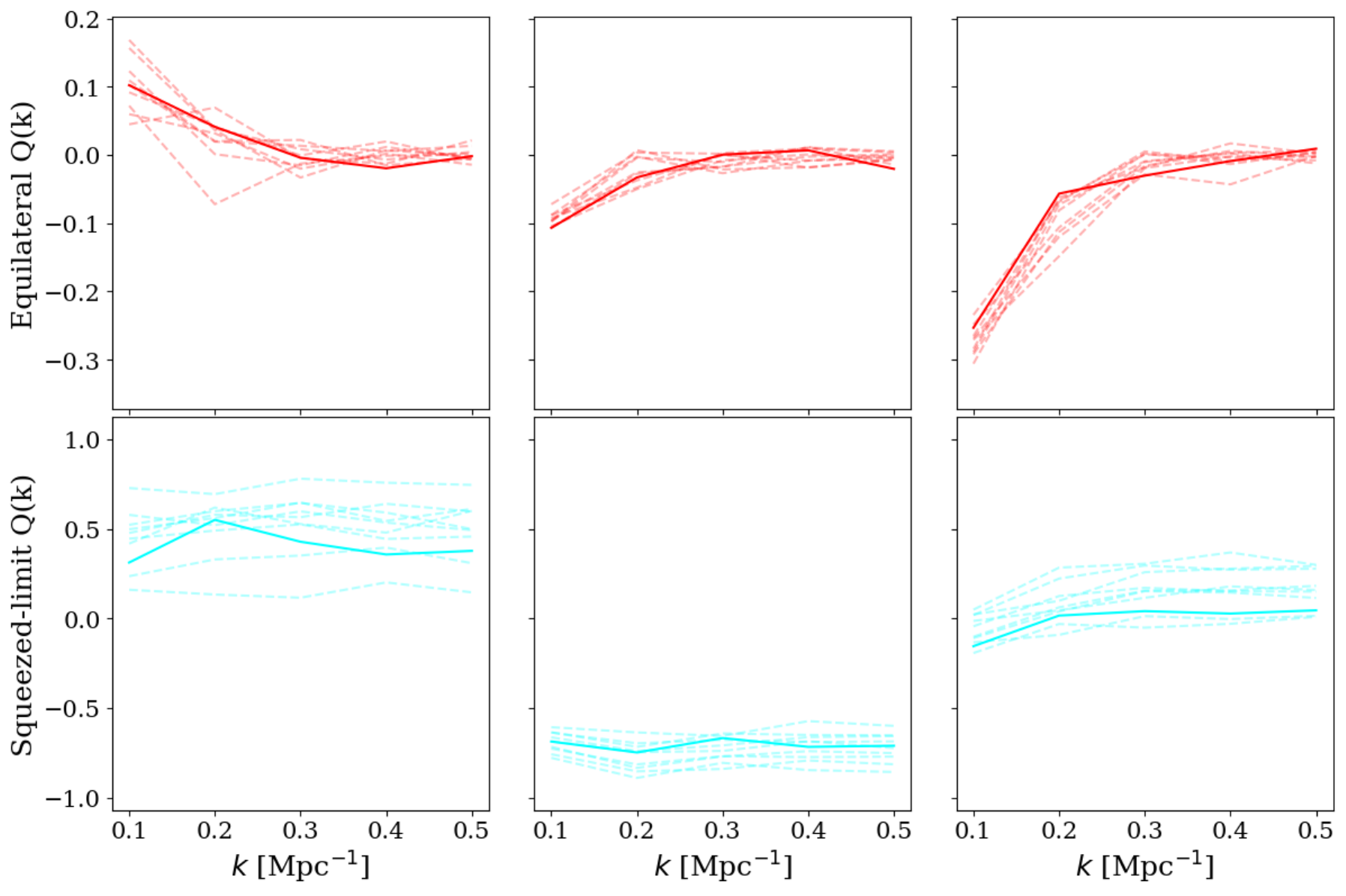}
    \includegraphics[width=0.80\textwidth, trim=0.4cm 0.5cm 0.4cm 0.4cm, clip]{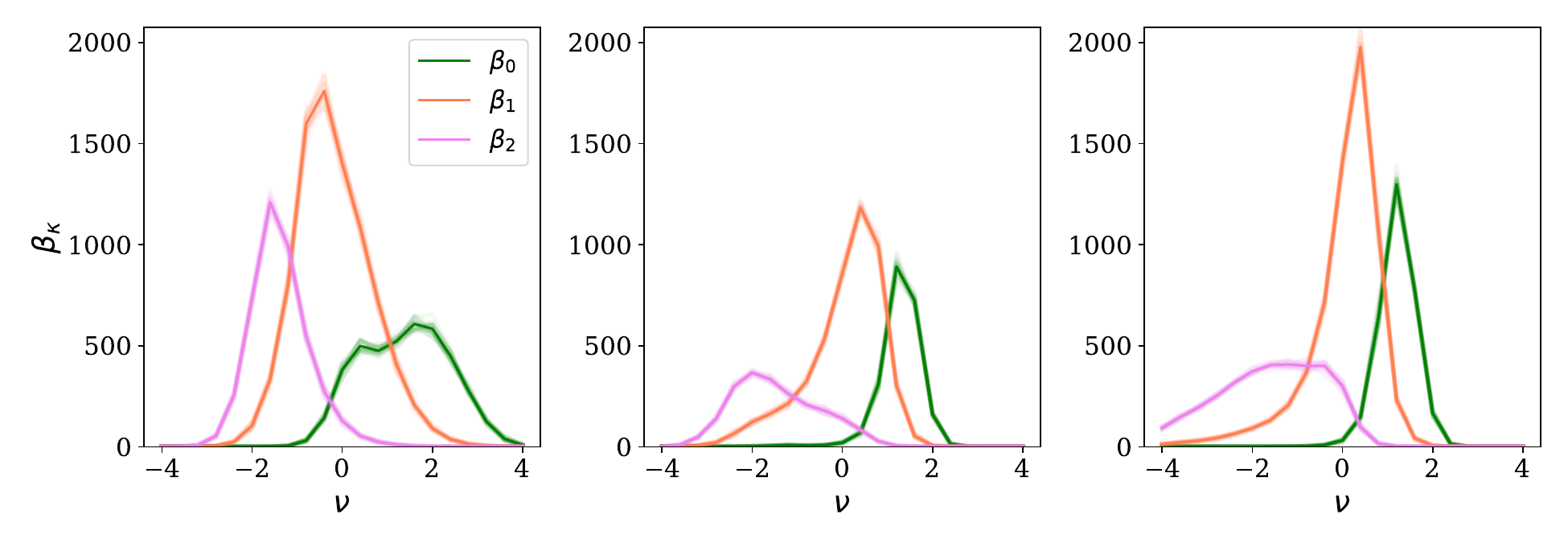}
    \caption{
    Non-Gaussian summary statistics at $z=7.2$ for the three reference models (Early, Fiducial, Late), assuming 1000h SKA-Low noise. \textit{Top row:} The reduced equilateral bispectrum, $Q(k)$. \textit{Middle row:} The reduced squeezed-limit bispectrum, $Q(k)$. \textit{Bottom row:} The Betti numbers ($\beta_0, \beta_1, \beta_2$) as a function of the threshold $v$. In all panels, light lines show multiple noise realisations, and the dark lines highlight a single, common realisation.
    }
    \label{fig:stat_examples_higherorder}
\end{figure*}

A direct (or field-level) comparison of the simulated 21-cm data with observations presents a high-dimensional inference problem that is computationally prohibitive. Furthermore, initial data from interferometers such as SKA-Low are expected to be noise-dominated, making such a direct analysis sub-optimal. We therefore compress this information by estimating several summary statistics from the cubes at different redshift bins, incorporating the instrumental effects described above. These statistics serve as the input data for our inference framework.

\subsubsection{2-point statistics}\label{sec:summary_gaussian}

The primary statistic in 21-cm cosmology to extract information is the power spectrum (PS), which quantifies the variance of the signal fluctuations as a function of spatial scale. As a two-point statistic, it fully describes a Gaussian random field and is proportional to $\bar{x}^2_{\rm HI}$ \citep[e.g.,][]{furlanetto2004growth,georgiev2022large}. We consider two versions. The first is the spherically averaged power spectrum (PS1D), $P_{\rm 1D}(k)$, which averages the signal power within spherical shells in Fourier space, providing information on the overall amplitude of fluctuations at different scales $k$. The second is the cylindrically averaged power spectrum (PS2D), $P_{\rm 2D}(k_{\perp}, k_{\parallel})$, which averages power in cylindrical annuli, separating scales perpendicular ($k_{\perp}$) and parallel ($k_{\parallel}$) to the line-of-sight. This retains information about line-of-sight effects such as redshift-space distortions \citep[e.g.,][]{jensen2013probing}. Both statistics are computed using modules available within \textsc{Tools21cm} \citep[][]{giri2020tools}.

Fig.~\ref{fig:stat_examples_gaussian} illustrates these Gaussian statistics at $z=7.2$ for our three reference models. The top row shows the PS1D for the three models. The different light solid lines in each panel show the PS1D of the signal with several different noise realisations for a 1000-hour SKA-Low observation, highlighting the variations introduced by instrumental effects. Instrumental noise adds a positive bias to the power spectrum, which can make the variations between realisations difficult to see. To better visualise these variations, we have removed this noise bias from the curves. One realisation is highlighted with a darker coloured curve. We can see that the shape and amplitude of the PS1D are distinct for each reference model, reflecting their different average neutral fractions at this redshift.

The bottom row of Fig.~\ref{fig:stat_examples_gaussian} shows the PS2D for the same single noise realisation as the dark curve in the top panel, but with the noise bias subtracted (similar to the PS1D plots). This bias subtraction reveals the underlying signal structure but also highlights the impact of noise. For instance, in the `Early' model, which has the lowest signal amplitude, many pixels are white. This is because these pixels were noise-dominated, and subtracting the large noise bias resulted in negative values, which are not displayed on the colour map. This illustrates the different noise properties of the two statistics. The PS1D benefits from more compression (averaging within spherical shells) and thus has a lower relative noise error. The PS2D, by preserving the $k_{\perp}$ and $k_{\parallel}$ separation, is more prone to noise but contains more astrophysical information from the signal's anisotropy. We will test and compare the physical information that can be extracted from both metrics in the results (Sec.~\ref{sec:results}). It is important to note that our inference framework does not subtract the noise bias; it is trained on and compares the statistics (signal + noise) directly. This avoids any issues related to bias subtraction, such as the negative values seen in the figure.

\subsubsection{Statistics sensitive to non-Gaussian information}\label{sec:summary_nongaussian}

The 21-cm signal is expected to be highly non-Gaussian, reflecting the complex morphology of ionised regions \citep[e.g.,][]{iliev2006simulating,giri2018bubble,giri2019position}. To capture this information, we compute the bispectrum, the three-point correlation function in Fourier space, $B(k_1, k_2, k_3)$, which is explicitly sensitive to non-Gaussianity. We use the publicly available \textsc{Pylians3} package \citep[][]{Pylians}, which implements a Fast Fourier Transform (FFT) based algorithm to estimate it \citep[e.g.,][]{Sefusatti2016accurate,Watkinson2017fast}. This algorithm requires the input data to be on a cubic grid. Since our lightcone data volumes are generally not cubic, we crop the data along the angular dimensions to create cubic sub-volumes before calculating the bispectrum.

The bispectrum depends on the configuration of the three wavevectors $\mathbf{k}_1, \mathbf{k}_2, \mathbf{k}_3$ (where $\mathbf{k}_1 + \mathbf{k}_2 + \mathbf{k}_3 = 0$). We focus on two specific configurations known to be sensitive probes of non-Gaussianity (see \citealt{lewis2011shape} for a detailed study of shapes probed by different configurations):
\begin{itemize}
    \item \textit{Equilateral}: $k_1 = k_2 = k_3 = k$. This configuration is sensitive to the typical shapes of the ionised regions and their clustering \citep[e.g.,][]{watkinson201921,majumdar2018quantifying}. 
    \item \textit{Squeezed-limit}: $k_1 \approx k_2 \gg k_3$. This configuration primarily probes the coupling between large-scale modulation and small-scale signal fluctuations \citep[e.g.,][]{giri2019position}. 
\end{itemize}
We assume a $k$-bin width of 0.1 Mpc$^{-1}$ for these calculations. Our analysis specifically uses the reduced bispectrum, which is defined as \citep[e.g.,][]{scoccimarro1998nonlinear}:
\begin{eqnarray}
    Q(k_1, k_2, k_3) = \frac{B(k_1, k_2, k_3)}{P(k_1)P(k_2) + P(k_2)P(k_3) + P(k_3)P(k_1)} \ .
\end{eqnarray}
This normalisation isolates the purely non-Gaussian part of the signal. The top two rows of Fig.~\ref{fig:stat_examples_higherorder} show this reduced bispectrum for the `Equilateral' and `Squeezed-limit' configurations, respectively. As in Fig.~\ref{fig:stat_examples_gaussian}, the columns correspond to the three reference models, and the plots show multiple noise realisations (faint lines) with one highlighted (dark curve). These plots show that the shape and amplitude of the reduced bispectrum also vary significantly between the models. Later in our forecast study, we infer constraints from both configurations jointly and refer to this combination as `Bispec'.

In contrast to the Fourier-space statistics mentioned above, we also compute Betti numbers. These are topological invariants calculated directly from the 21-cm signal images, which are an expected data product from SKA-Low \citep[e.g.,][]{mellema2015hi,giri2021measuring}. Specifically, $\beta_0$ counts the number of connected components (e.g., isolated ionised regions), $\beta_1$ counts the number of tunnels or loops, and $\beta_2$ counts the number of enclosed voids. We calculate these as a function of a threshold value $v$, defined relative to the standard deviation of the data ($v = \delta T_b / \sigma_{\delta T_b}$), using 21 threshold bins between $v = -4$ and $4$. This provides information about the topology at different intensity levels\footnote{This thresholding strategy or \textit{filtration} differs from that used in \citet{giri2021measuring} where ionised regions were first explicitly identified using a pattern recognition framework across all redshifts, and Betti numbers were then calculated as a function of redshift. Our approach focuses on topology as a function of signal intensity within the three specific redshift bins considered in this study.}.

The instrumental noise in the images is approximately Gaussian in nature \citep[e.g.,][]{giri2018optimal}. For a reference Gaussian random field, the Betti number curves are known to be symmetric, with $\beta_0$ and $\beta_2$ as mirror images of each other \citep[e.g.,][]{pranav2019topology,giri2021measuring}. The bottom row of Fig.~\ref{fig:stat_examples_higherorder}, showing the Betti curves for our reference models, clearly deviates from this symmetry. The distinct, asymmetric shapes of these curves, particularly the peaks and their locations, demonstrate that they capture the unique non-Gaussian topological signatures of each reionization scenario, which are invisible to the power spectrum.

\section{Implicit likelihood and prior inference}\label{sec:inference}
We perform the joint inference of $\mathbf{\bar{x}}_{\rm HI}=\left[\bar{x}_{\rm HI}(z=8.0), \bar{x}_{\rm HI}(z=7.2), \bar{x}_{\rm HI}(z=6.5)\right]$ given $\mathbf{t}$, a summary statistic measured in the three redshift (or frequency) bins of the study (see \S\ref{sec:summary_stats}).
Our data model provides samples $(\mathbf{\bar{x}}_{\rm HI}, \mathbf{t})$ drawn from the simulator, which can be seen as an implicit likelihood distribution $p(\mathbf{\bar{x}}_{\rm HI}, \mathbf{t} | \mathbf{\theta})$. We aim to estimate the posterior distribution $p(\mathbf{\bar{x}}_{\rm HI} | \mathbf{t})$, which following Bayes' theorem can be evaluated as: 
\begin{equation}\label{eq:bayes}
    p(\mathbf{\bar{x}}_{\rm HI} | \mathbf{t}) \propto p(\mathbf{t} | \mathbf{\bar{x}}_{\rm HI}) ~ p(\mathbf{\bar{x}}_{\rm HI}).
\end{equation}
\par However, unlike in a classical forward model where the inferred quantity is an input to the model, here the mean neutral fraction $\mathbf{\bar{x}}_{\rm HI}$ is one of its output. Hence, the simulator carries an implicit prior $p(\mathbf{\bar{x}}_{\rm HI})$, which can be written as a function of the simulator's implicit likelihood and the simulator's prior $p(\theta)$ (see \S\ref{sec:sampling_prior}):
\begin{equation}
    p(\mathbf{\bar{x}}_{\rm HI}) = \iint p(\mathbf{\bar{x}}_{\rm HI}, \mathbf{\mathbf{t}}| \mathbf{\theta}) p(\mathbf{\theta})\rm{d}\mathbf{\theta}\rm{d}t.
\end{equation}
Since we do not have explicit expressions for both $p(\mathbf{\bar{x}}_{\rm HI}, \mathbf{\mathbf{t}}| \mathbf{\theta})$ and $p(\mathbf{\theta})$, it is clear that the prior $p(\mathbf{\bar{x}}_{\rm HI})$ is implicit. Likewise, the likelihood $p(\mathbf{t} | \mathbf{\bar{x}}_{\rm HI})$ in eq. \ref{eq:bayes} can be written as a function of the simulator's implicit likelihood and prior:
\begin{align}
    p(\mathbf{t} | \mathbf{\bar{x}}_{\rm HI}) &= \int  
    \frac{p(\mathbf{\bar{x}}_{\rm HI}, \mathbf{\mathbf{t}}| \mathbf{\theta})}{p(\mathbf{\bar{x}}_{\rm HI})}p(\theta)\rm{d}\mathbf{\theta}.
\end{align}
As all terms are intractable in this expression, it demonstrates that also the likelihood term of in Eq.~(\ref{eq:bayes}) is implicit. While implicit inference is classically required when the likelihood only is intractable, it is in our case also justified by the intractable prior on $\mathbf{\bar{x}}_{\rm HI}$. This problem requires us to use Neural Posterior Estimation \citep[NPE,][]{papamakarios2018NPE}, as it is the only class of implicit inference algorithm that learns the prior together with the likelihood. 

Here, we follow a state-of-the-art inference strategy to model $p(\mathbf{\bar{x}}_{\rm HI} | \mathbf{t})$.  We use the NPE method, e.g. training a conditional distribution $q_\varphi(\mathbf{\bar{x}}_{\rm HI}|\cdot)$ with weights $\varphi$ to learn both the likelihood and prior. Instead of conditioning $q_\varphi$ directly on the summaries $\mathbf{t}$, we perform variational mutual information maximisation \citep[VMIM,][]{JeffreyVMIM2020} to optimally extract the relevant information from $\mathbf{t}$. To do so, we train a neural network $\mathbf{y}_{\psi}(\mathbf{t})$ with weights $\psi$. Its output yields the same shape as $\mathbf{\bar{x}}_{\rm HI}$. In practice, $\mathbf{y}_{\psi}(\mathbf{t})$ is obtained by training $\mathbf{y}_{\psi}$ simultaneously with $q_\varphi$. Hence, we minimise the combined VMIM loss:

\begin{equation}
    L_{\rm VMIM}(\varphi, \psi) = -\mathbb{E}_{p(\mathbf{x}_{\rm HI},\mathbf{t})}{\log(q_\varphi(\mathbf{x}_{\rm HI} | \mathbf{y}_{\psi}(\mathbf{t})))}.
\end{equation}
For $\mathbf{y}_{\psi}(\mathbf{t})$, we opt for a simple network architecture that can be indifferently applied to all data types. This allows for a fair comparison of the posterior estimators. We used a fully connected architecture with 3 dense layers of 256 neurons and Rectified Linear Unit activations. In \cite{de2025exploring} we explored complex deep neural architectures to infer $\mathbf{\bar{x}}_{\rm HI}$ from PS2D; we devote a similar study for higher-order statistics to a future work. For $q_\varphi$, we use the regular NPE implentation in the \textsc{sbi} package \citep[][]{tejero-cantero2020sbi}. For an optimal compression, $p(\mathbf{\bar{x}}_{\rm HI} | \mathbf{y})$ should be identical to $p(\mathbf{\bar{x}}_{\rm HI} | \mathbf{t})$. As such, we refer similarly to both distributions as `the posterior'. 

When combining different kind of statistics $(\mathbf{t}_1, \dots , \mathbf{t}_n)$, we separately train the VMIM compressors and concatenate their respective outputs ($\mathbf{y}_1, \dots, \mathbf{y}_n)$ into $\mathbf{y}_{1\times\dots\times n}$. Then, the NPE training is performed on $\mathbf{y}_{1\times\dots\times n}$. In the following, we will focus on combinations of gaussian and higher-order statistics, in particular PS2D+Betti, PS2D+Bispec, and PS2D+Betti+Bispec. 

Through the simulation of $\sim15,900$ independent mock lightcones and the various noise realisations applied to each (see \S\ref{sec:sampling_prior} and \S\ref{sec:instrumental_effects}), we end up with a sample of $\sim$330,000 $(\mathbf{\bar{x}}_{\rm HI}, \mathbf{t})$ pairs, of which 170,000 (30,000) are used for the VMIM training (validation), 60,000 (20,000) for the NPE training (validation), and the remaining for testing. To check the robustness of our approach, we train 30 independent compressors and posteriors for each level of noise and kind of summary $\mathbf{t}$.

\section{Results}\label{sec:results}

\subsection{Inference framework calibration}\label{sec:results_calibration} % \subsection{Metrics and calibration}

\begin{figure}
    \centering
    \includegraphics[width=\linewidth, trim=0.9cm 1.0cm 1.5cm 1.5cm, clip]{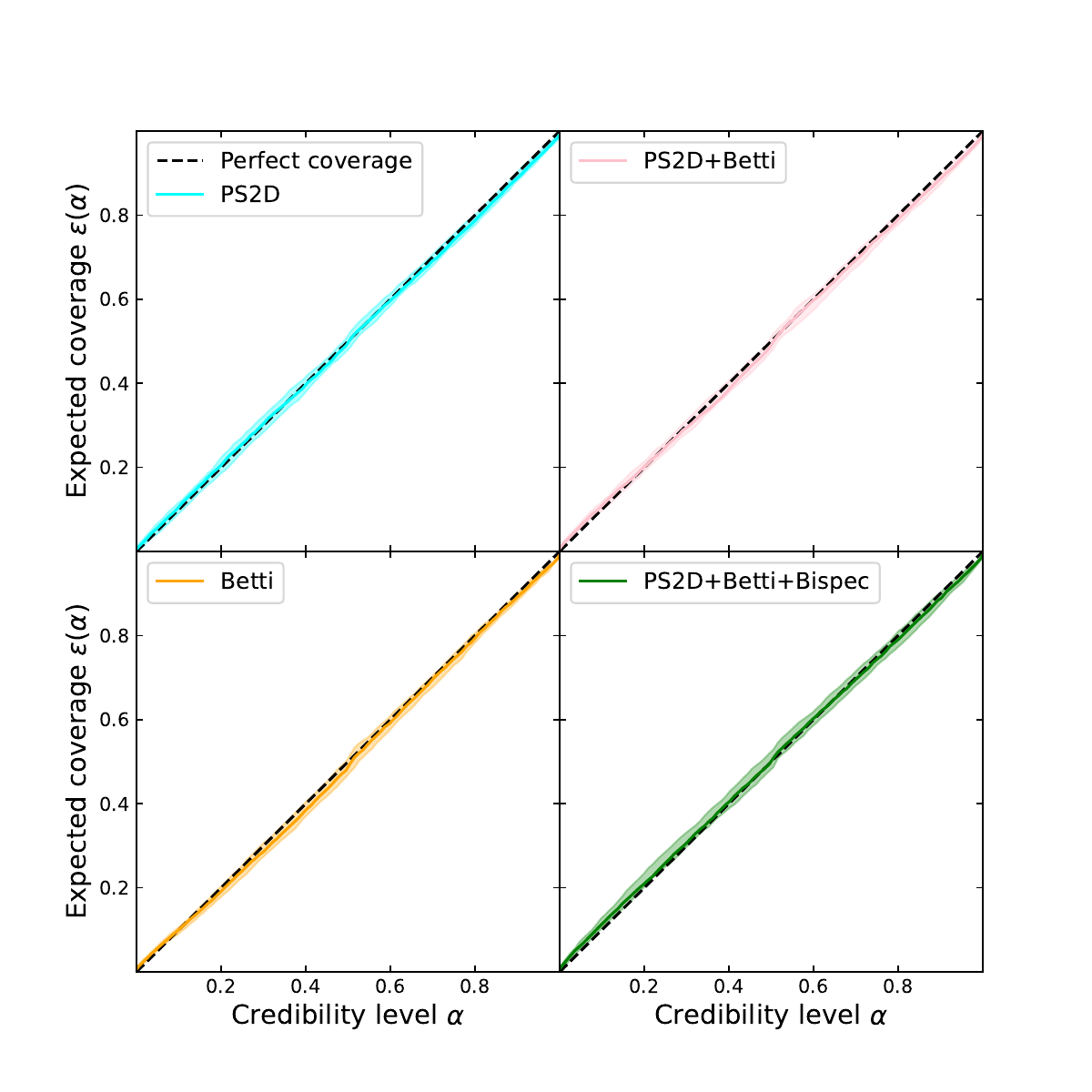}
    \caption{Coverage test on the best-calibrated models for PS2D (cyan), Betti numbers (gold), PS2D+Betti (pink) and PS2D+Betti+Bispec (green). For each model we ran 20 TARP realisations and show medians as solid lines and 95\% of the samples as shaded regions.}
    \label{fig:tarps}
\end{figure}

\begin{figure*}
    \centering
    \includegraphics[width=0.7\linewidth]{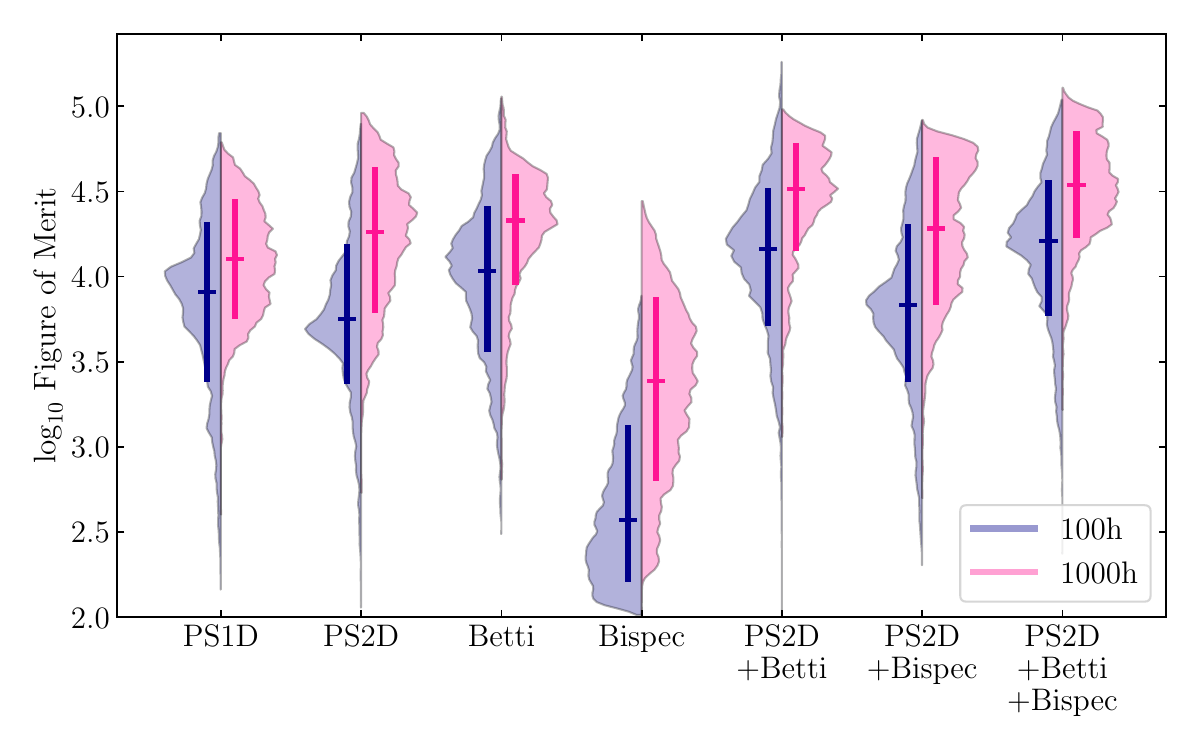}
    \caption{Figure of merit distributions for all tested summary statistics, from left to right: spherical power spectrum (PS1D), cylindrical power spectrum (PS2D), Betti numbers (Betti), reduced bispectrum (Bispec), and their combinations PS2D+Betti, PS2D+Bispec, PS2D+Betti+Bispec. Horizontal bars denote the median of the distribution while vertical bars range from the 16th to the 84th percentiles of the samples. We show the results for both 100h (blue) and 1000h (pink) SKA-Low AA* noise level.
    }
    \label{fig:FoMdistribs}
\end{figure*}

\begin{table*}
\centering
\caption{Figure of merit comparison. The central part of the table shows $\log_{10}$ FoM for sets of summary statistics (single and combined). We quote for 100h (top section) and 1000h (central section) SKA-Low AA* noise level, the 16th, 50th 84th percentiles of the FoM distributions sampled over 4000 test mocks. The right-hand side of the table shows the relative improvement of these quantities going from PS2D alone to PS2D+Betti and PS2D+Betti+Bispec. The bottom section of the table shows the relative change of the 50th percentile going from 100h to 1000h noise level.} \label{tab:FoMresults}

\begin{tabular}{ll|ccccccc|cc}
\toprule &  & \multicolumn{7}{c}{$\log_{10}$ FoM} & \multicolumn{2}{c}{Relative improvement (dex)} \\ \makecell{noise\\level} & \makecell{percen-\\tiles} & PS1D & PS2D & Betti & Bispec & \makecell{PS2D\\+Betti} & \makecell{PS2D\\+Bispec} & \makecell{PS2D\\+Betti\\+Bispec} & \makecell{(PS2D+Betti)\\/PS2D} & \makecell{(PS2D+Betti\\+Bispec)/PS2D}\\
\midrule 
\multirow[t]{3}{*}{100h} & 16th & 3.40 & 3.39 & 3.58 & 2.22 & 3.73 & 3.40 & 3.79 & 0.34 & 0.40 \\
& 50th & 3.91 & 3.75 & 4.03 & 2.57 & 4.16 & 3.83 & 4.21 & 0.41 & 0.46 \\
& 84th & 4.30 & 4.17 & 4.40 & 3.11 & 4.50 & 4.29 & 4.55 & 0.33 & 0.37 \\
\cline{1-11}
\multirow[t]{3}{*}{1000h} & 16th & 3.77 & 3.80 & 3.97 & 2.82 & 4.17 & 3.85 & 4.25 & 0.37 & 0.44 \\
& 50th & 4.10 & 4.26 & 4.33 & 3.39 & 4.51 & 4.28 & 4.54 & 0.25 & 0.27 \\
& 84th & 4.44 & 4.63 & 4.58 & 3.86 & 4.77 & 4.69 & 4.84 & 0.14 & 0.21 \\
\cline{1-11}
\makecell{1000h\\ / 100h \\ (dex)} & 50th & 0.19 & 0.51 & 0.29 & 0.81 & 0.35 & 0.45 & 0.33 & &  \\
\cline{1-11}
\bottomrule
\end{tabular}

\end{table*}

Assessing the calibration of the learnt posterior is critical for implicit inference methods. We here use the test of accuracy with random points \citep[TARP,][]{lemos2023TARP}, an efficient method to detect miscalibrated multidimensional posteriors. It provides us with the expected coverage probability as a function of the credibility level, $\varepsilon(\alpha)$. We computed the empirical frequency $\hat{p}_k$ of $\varepsilon(\alpha)$ over $K=100$ credibility level bins by bootstraping the TARP 40 times. Similarly to \cite{Saoulis_2025}, we then computed the calibration error as the mean squared error between $\hat{p}_k$ and the ideal $p_k$:
\begin{equation}
    \mathcal{C} = \frac{1}{K}\sum_k 
    \left( \frac{\hat{p}_k - p_k}{p_k}
    \right)^2.
\end{equation}
For an ideally calibrated posterior, $\varepsilon(\alpha)=\alpha$ and hence $\hat{p}_k = p_k = 1/K$. We computed the calibration error for all learnt posterior, and selected for each summary (including combinations of summaries) the posterior model that achieves the lower $\mathcal{C}$. These best-calibrated posteriors typically yield $\mathcal{C} \sim 0.002$. In Fig. \ref{fig:tarps} we show the TARP outputs $\varepsilon(\alpha)$ for PS2D (cyan), Betti numbers (gold), PS2D+Betti (pink) and PS2D+Betti+Bispec (green). Solid lines indicate the median from 20 TARP realisations and shaded regions encompass 95\% of the samples. From the figure we do not note any significant deviation from the 1:1 line, meaning that these posteriors are unbiased and well calibrated. In the remaining of the paper, when referring to a posterior estimator of a specific kind, we in fact refer to the best-calibrated model for the corresponding summary.

To estimate the amount of information provided by a posterior, we employ the figure of merit (FoM). We empirically estimate the posterior covariance Cov$[\mathbf{\theta}|\mathbf{t_0}]$ for a test $\mathbf{t_0}$ and obtain the FoM:
\begin{equation}
    \text{FoM} = \left| \text{Cov}[\mathbf{\theta} |\mathbf{t_0}] 
    \right|^{-1/N},
\end{equation}
with $N$ the number of dimension of the posterior, 3 in our case, and $\left|\cdot \right |$ the matrix determinant. Since the determinant scales as the $N$-th power of the covariance matrix elements, in our definition the FoM roughly scales as the inverse square of the standard deviation of the posterior samples, or $\sigma^{-2}$. For instance, an increase of the FoM of 0.3 dex represents an decrease of $\sim 40\%$ on the uncertainty on \barxHI{}. For each model, we selected 4000 random $\mathbf{t_0}$ samples from the test set and measured the FoM of their associated posterior.

\subsection{Comparison of constraining power of the statistics}\label{sec:results_constraints} %\subsection{Statistic comparison}

\begin{figure*}
    \centering
    \includegraphics[width=0.49\linewidth]{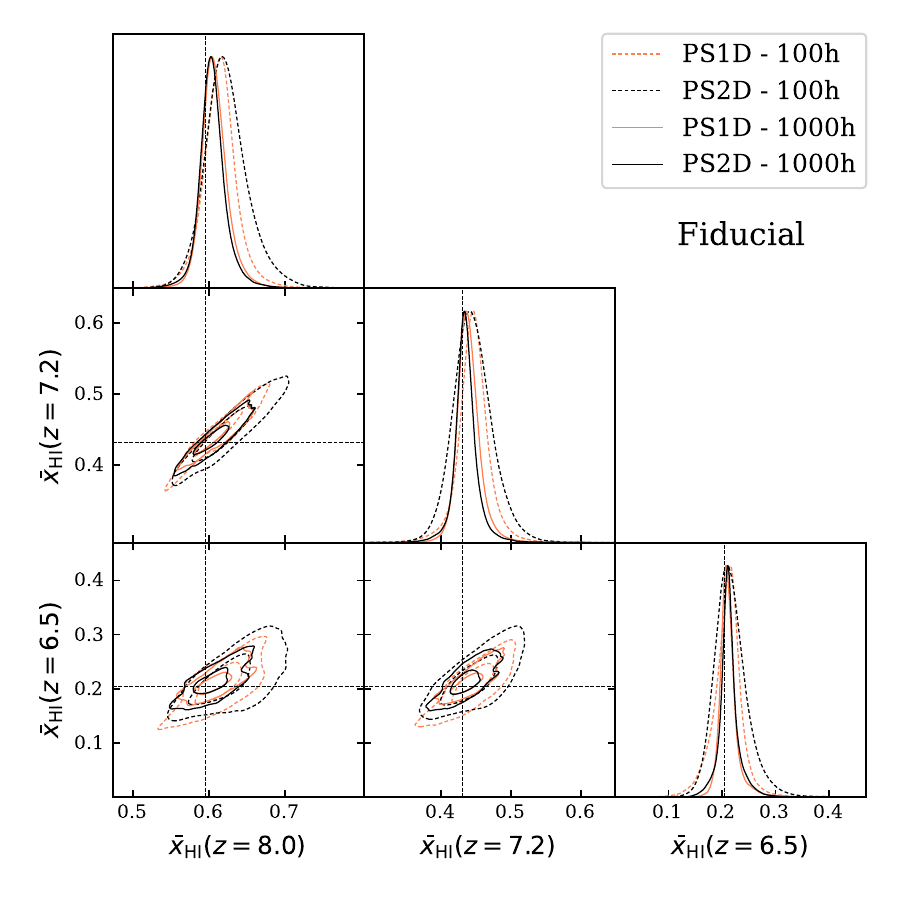}
    \includegraphics[width=0.49\linewidth]{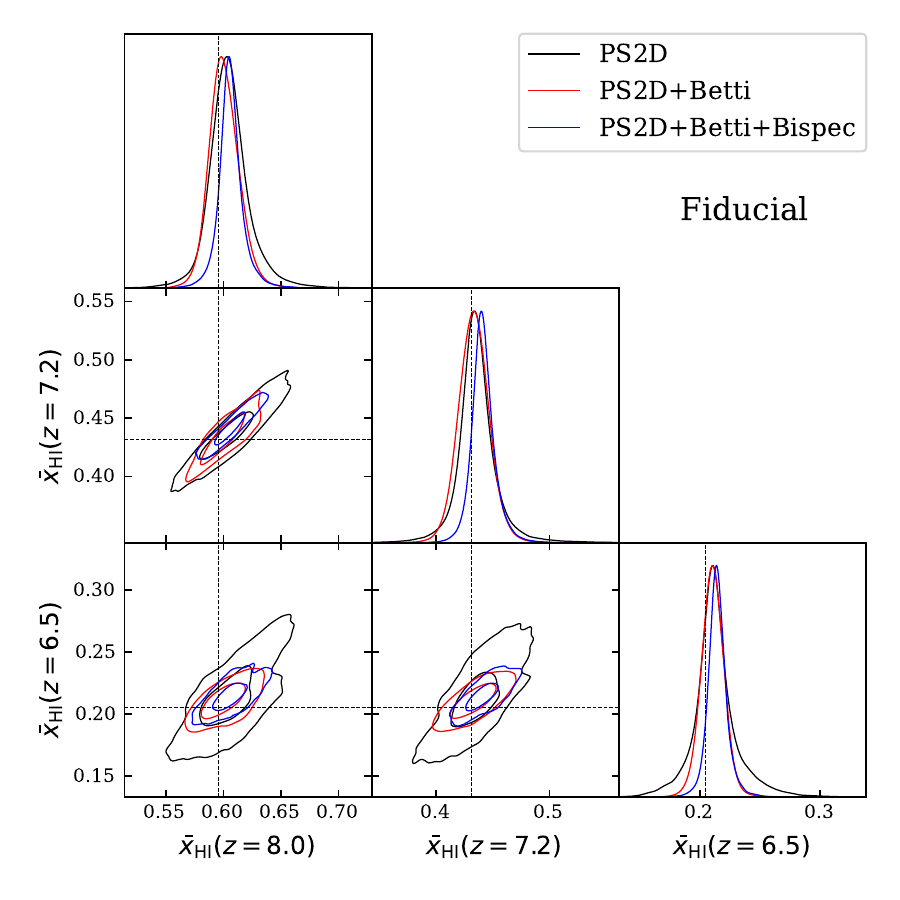}
    \includegraphics[width=0.49\linewidth]{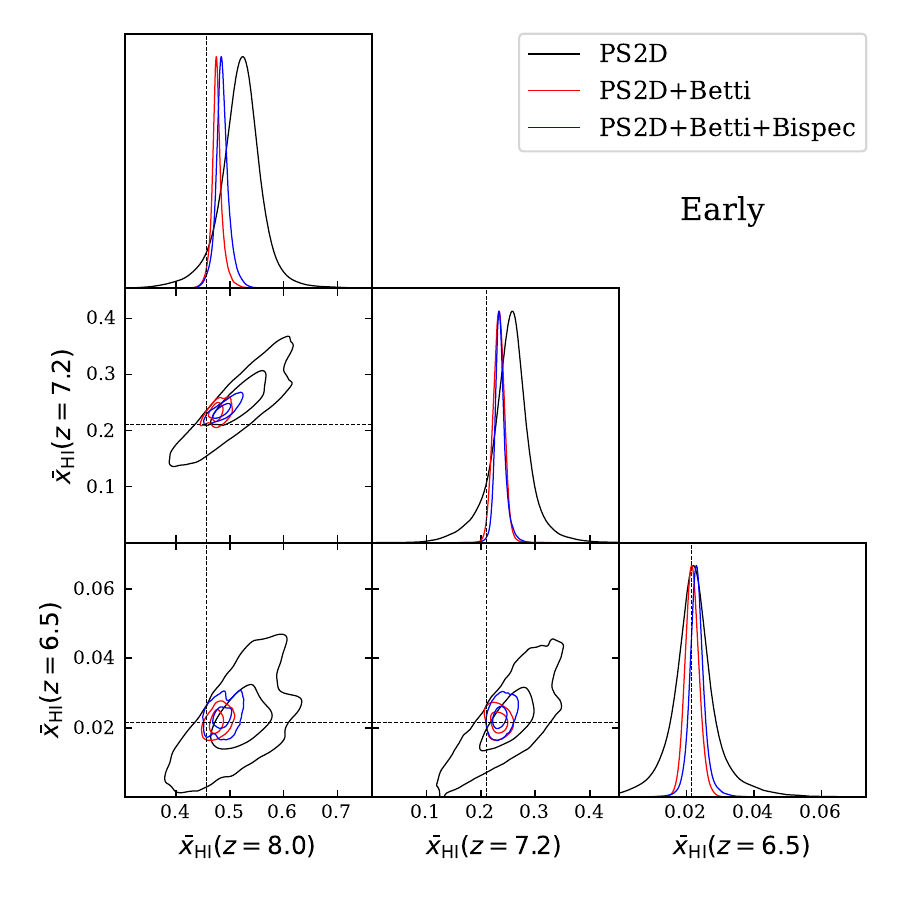}
    \includegraphics[width=0.49\linewidth]{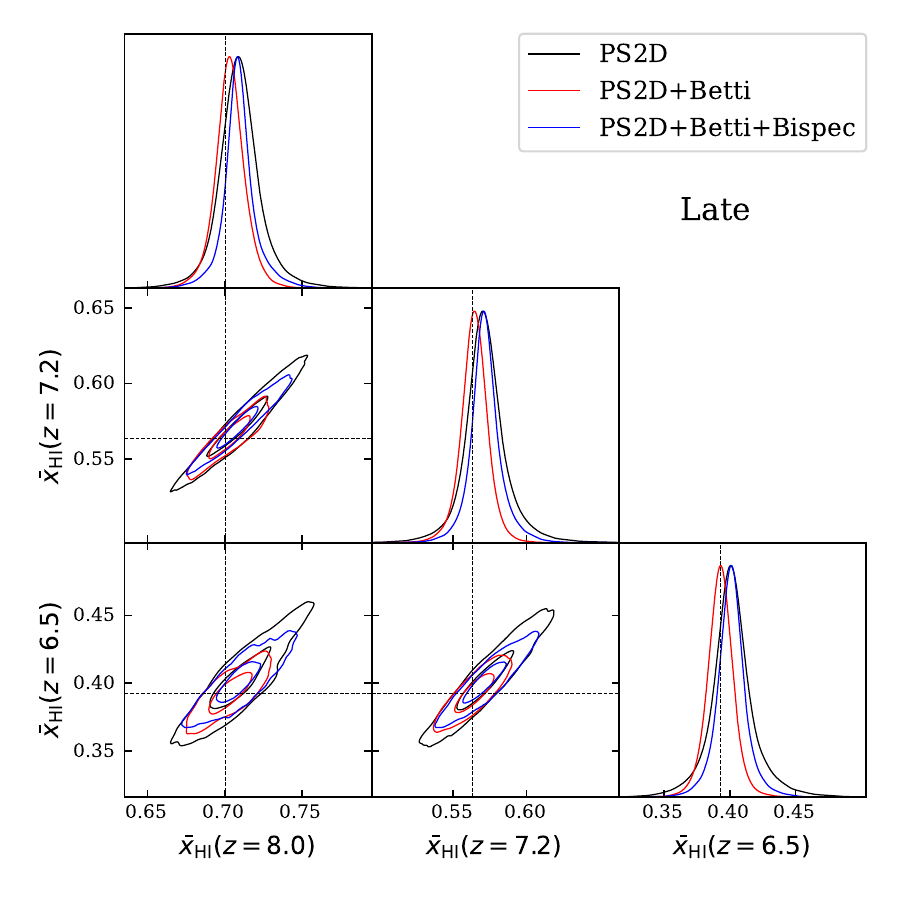}
    \caption{
    Posterior constraints on the 3D reionization history ($\bar{x}_{\rm HI}(z=8.0)$, $\bar{x}_{\rm HI}(z=7.2)$, $\bar{x}_{\rm HI}(z=6.5)$) for our three reference mocks. \textit{Top-left}: Comparison of posteriors from PS1D and PS2D for the Fiducial mock, shown for both 100h (dashed lines) and 1000h (solid lines) SKA-Low noise. \textit{Other panels}: Comparison of posteriors from PS2D (black), PS2D+Betti (red), and PS2D+Betti+Bispec (blue), all assuming 1000h noise. The mock is the Fiducial model (top-right), the Early model (bottom-left), and the Late model (bottom-right). The true values for each mock are indicated by the dashed grey lines.
    }
    \label{fig:contours_all}
\end{figure*}

\begin{table*}
\centering
\caption{Constraints on reionization history for the three reference scenarios: Fiducial, Early and Late; at the central redshifts of the simulated boxes. We vary the statistics used (PS1D, PS2D, PS2D+Betti or PS2D+Betti+Bispec) and the level of noise (100h or 1000h). We quote the median and 68\% confidence interval limits. } \label{tab:xHIconstraints}
\renewcommand{\arraystretch}{1.2}
\begin{tabular}{lll|ccc}

\toprule 
Scenario & Statistics & Noise level & $x_{\rm HI}(z=8.0)$ & $x_{\rm HI}(z=7.2)$ & $x_{\rm HI}(z=6.5)$ \\
\midrule 
\multirow{6}{*}{Fiducial} & PS1D & 100h & $0.614\pm 0.025$ & $0.446\pm 0.026 $ & $0.210^{+0.025}_{-0.022}   $ \\
                          & PS2D & 100h & $0.623^{+0.023}_{-0.031}$ & $0.446^{+0.024}_{-0.030}$ & $0.217^{+0.023}_{-0.034}$\\
                          & PS1D  & 1000h & $0.608^{+0.013}_{-0.019}$ & $0.440^{+0.012}_{-0.017}$ & $0.213^{+0.009}_{-0.017}$\\
                          & PS2D  & 1000h & $0.605^{+0.013}_{-0.017}$ & $0.436^{+0.012}_{-0.014}$ & $0.213^{+0.013}_{-0.016}$ \\
                          & PS2D + Betti & 1000h & $0.601^{+0.011}_{-0.014}$ & $0.433\pm 0.014            $ & $0.211\pm 0.010$\\
                          & PS2D + Betti + Bispec & 1000h & $0.606^{+0.008}_{-0.010} $ & $0.442^{+0.008}_{-0.010} $  & $0.215^{+0.007}_{-0.008}$\\
\cline{1-6}
\multirow{3}{*}{Early} & PS2D & 1000h & $0.520^{+0.037}_{-0.032}$ & $0.253^{+0.034}_{-0.029}$ & $0.023^{+0.005}_{-0.006}$ \\
                       & PS2D + Betti & 1000h & $0.477^{+0.008}_{-0.010}$ & $0.233\pm 0.011$ & $0.022^{+0.002}_{-0.002}$\\
                       & PS2D + Betti + Bispec & 1000h & $0.486^{+0.010}_{-0.012} $ & $0.236^{+0.007}_{-0.010} $ & $0.023^{+0.002}_{-0.002}$ \\
\cline{1-6}
\multirow{3}{*}{Late} & PS2D & 1000h & $0.709^{+0.012}_{-0.013}   $ & $0.572^{+0.012}_{-0.013}   $ & $0.402\pm 0.018            $\\
                      & PS2D + Betti & 1000h & $0.704\pm 0.011$  & $0.564\pm 0.010$ & $0.393\pm 0.011$ \\
                      & PS2D + Betti + Bispec & 1000h & $0.709^{+0.008}_{-0.010}$  & $0.572^{+0.009}_{-0.010}$ & $0.401\pm 0.013$ \\
\cline{1-6}
\bottomrule
\end{tabular}
\end{table*}

We start by comparing the FoM distributions for all best-calibrated models. In Fig. \ref{fig:FoMdistribs} we show from left to right the FoM distributions from PS1D, PS2D, Betti, Bispec, PS2D+Betti, PS2D+Bispec and PS2D+Betti+Bispec. The distributions are represented in both cases of noise level, 100h (blue) and 1000h (pink). To help comparing the distributions, we show the median (horizontal bar) and the [16th-84th] percentiles interval (vertical bar). We also provide the values of the median, 16th and 84th percentiles for all distributions in table \ref{tab:FoMresults}.  As expected, all distributions systematically shift upwards when the observational time is ten times longer. The PS1D shows the mildest improvement (0.19 dex) from 100h to 1000h, while PS2D yields a 0.51 dex increase in FoM. The bispectrum has the biggest increase (0.81 dex), but remains significantly less informative than other statistics.
\par For gaussian statistics alone, we observe that the PS1D and the PS2D have roughly equivalent FoMs. At 100h, the PS2D is slightly below the PS1D, but we observe an opposite trend at 1000h. This shows that the PS2D is more informative than the PS1D, provided that there is sufficient signal-to-noise. While quantifying the SNR regime where the PS2D overtakes the PS1D would be useful, we expect it to be dependent on the analysis setup. We hence devote the study of this transition to future work, with more advanced modelling (e.g. including foreground residuals and antenna calibration errors).
\par Then, looking at the higher-order statistics alone, the Betti numbers appears to be more informative than the PS2D. However, this is not true for the Bispec, which FoMs are about 1.5 dex below the other summaries alone. Several reasons may explain this. First, the bispectrum computation required an angular scale cut, excluding the largest scales that are the most informative on $\bar{x}_{\rm HI}$, as explained in \S\ref{sec:summary_nongaussian}. Then, we computed the reduced bispectrum, in which 2pt-information are cancelled out. 
\par Lastly, we look at the FoMs for combined statistics. The PS2D+Betti median is improved by 0.25 (0.18) dex with respect to PS2D (Betti) alone. These summaries hence contain complementary information on \barxHI. The PS2D+Bispec also slightly improves on PS2D alone (0.02 dex for the median), even if the bispectrum alone is much less constraining \barxHI. This is in fact not unexpected as the reduced bispectrum captures only non-gaussian information, while the PS2D is restriced to gaussian information. There combination can only improve the quantity of information available on \barxHI. Finally, the combination PS2D+Betti+Bispec appears to be the most informative one. The median FoM is improved by 0.27 dex with respect to the PS2D alone. Interestingly, the combination of statistics benefits more to the lower end of the distribution (the 16th percentile is increased by 0.37 and 0.44 dex respectively for PS2D+Betti and PS2D+Betti+Bispec). 
\par The weak contribution of the bispectrum was also found in other studies. \cite{watkinson2022epoch} inferred the simulation parameters using the isosceles triangle configuration bispectrum, and found that, while it may provide complementary constraints to the power spectrum in an idealised case, its information is severly reduced when sample variance and SKA-Low instrumental noise are introduced in the analysis. Recently, \cite{krishna2026constrainingneutralhydrogenfraction} inferred \barxHI{} using the power spectrum combined with the full and not reduced bispectrum, and also found a mild contribution from the bispectrum in the presence of instrumental noise. Our findings are hence consistent with these works. We also recall that the bispectrum is the statistic with the strongest increase of FoM from 100h to 1000h (0.81dex), suggesting that the noise is a key factor to unlock the bispectrum information.

\subsection{Posterior constraints for reference scenarios}\label{sec:results_refposterior}

\begin{figure*}
    \centering
    \includegraphics[width=0.95\linewidth]{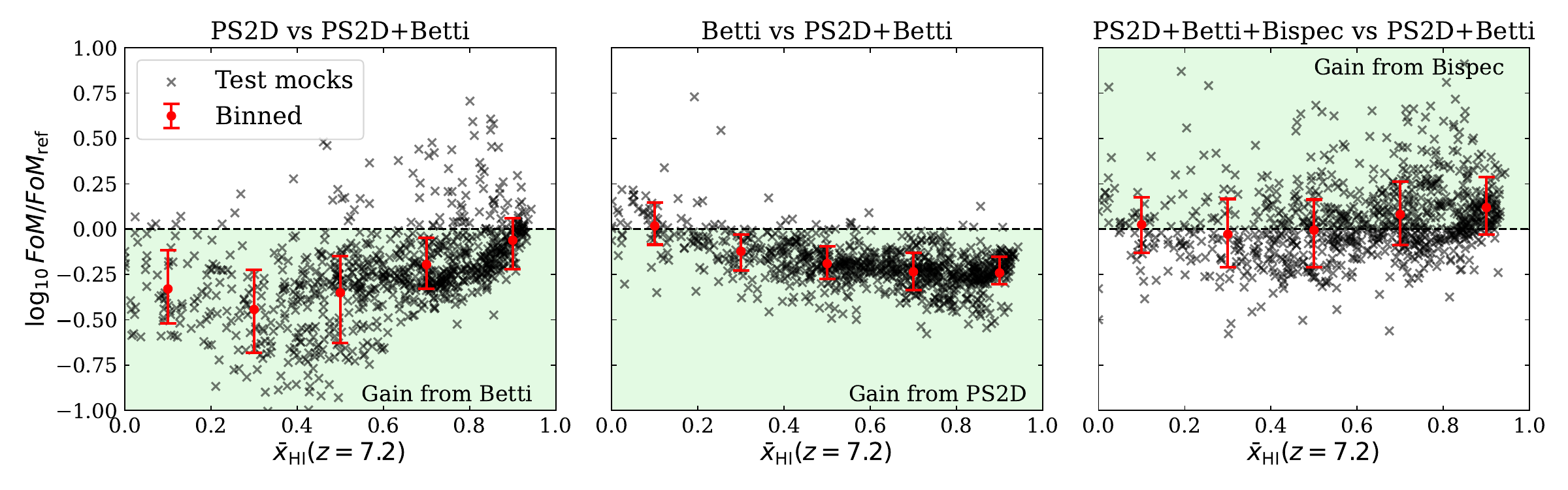}
    \caption{Relative improvement of the figure of merit as a function of the \xHI{} value in the central frequency range, for the 1000h noise level case. The reference posterior is PS2D+Betti across all panels: this makes all point of comparisons only one step away from the reference (i.e., one statistic is either removed or added). From left to right, we show the relative FoM of the posteriors from PS2D, Betti, and PS2D+Betti+Bispec. The ratio FoM/FoM$_{\rm ref}$ for each test mock is represented as a function of $x_{\rm HI}(z=7.2)$ (black crosses). We divide the data points into five \barxHI{} bins and show the median (horizontal red bar) and the (16th, 84th) percentiles interval (vertical red bar). In green we show the region where the additional summary leads to a gain in constraining power.}
    \label{fig:FoMratio}
\end{figure*}

Having compared the overall FoM distributions, we now examine the detailed 2D and 1D posterior constraints for our reference mocks, presented in Fig.~\ref{fig:contours_all}, along with the corresponding constraints on \barxHI{} in Table~\ref{tab:xHIconstraints}. The top-left panel of Fig.~\ref{fig:contours_all} focuses on the Fiducial mock, comparing the posteriors obtained from the PS1D and PS2D statistics at both 100h (dashed lines) and 1000h (solid lines) noise levels. As expected, the deeper 1000h observation yields significantly tighter constraints in all cases. We find that at 1000h, the constraining power of PS1D and PS2D is comparable. However, for the shallower 100h observation, the PS1D provides slightly better constraints than the PS2D, which is consistent with the FoM distributions for lower signal-to-noise ratio data we saw in Fig.~\ref{fig:FoMdistribs}.

The remaining three panels investigate the gain from adding non-Gaussian statistics, assuming the deep 1000h SKA-Low observation. Since PS1D and PS2D are comparable at this noise level, we use PS2D as the fiducial Gaussian baseline (black). The top-right panel shows these results for the Fiducial mock, the bottom-left panel for the Early mock, and the bottom-right panel for the Late mock. Each of these plots compares the constraints from PS2D (black) against those from PS2D+Betti (red) and the full combination of PS2D+Betti+Bispec (blue). In general, adding non-Gaussian information improves the constraints. For both the Fiducial and Early models, the posteriors shrink significantly when Betti numbers are included; in the case of the Early model they shrink further with the addition of the bispectrum. This demonstrates the complementary nature of these statistics. For completeness, we provide in Appendix \ref{app:hosalone} a comparison of all the individual statistics on the Fiducial model.

However, we observe an interesting exception in the Late model case (bottom-right). While adding Betti numbers (red) still provides a substantial improvement over PS2D alone, the full combination including the bispectrum (blue) results in a weaker constraint, with a posterior volume larger than that from PS2D+Betti alone. This suggests that the constraining power of a given statistic may not be uniform, but rather depends on the underlying reionization history (i.e., the neutral fraction). We explore this dependence in detail in the next section.

\subsection{Dependence of constraining power on reionization state}\label{sec:results_reion_dependence}

The results from the Late model in Fig.~\ref{fig:contours_all}, where adding the bispectrum to the PS2D+Betti combination unexpectedly degraded the posterior, suggest that the constraining power of different statistics is not uniform. This motivates a more detailed investigation into how the reionization state, i.e., the average neutral fraction $\bar{x}_{\rm HI}$, impacts the relative performance of our summary statistics. To explore this, we use our full test set of mocks. Fig.~\ref{fig:FoMratio} plots the relative FoM for the 1000h noise case as a function of the neutral fraction at $z=7.2$. We use the PS2D+Betti combination as a reference ($\rm{FoM}_{\rm{ref}}$) and plot the log-ratio of the FoM for other statistics relative to it. The faint crosses show individual mocks, while the red bars show the binned median and [16th-84th] percentile range.

The panels in Fig.~\ref{fig:FoMratio} reveal the complementary nature of the statistics and their state-dependence. The left panel shows the performance of PS2D relative to PS2D+Betti. The ratio is on average well below zero (log$_{10}({\rm FoM}/{\rm FoM}_{\rm ref}) < 0$), confirming that Betti numbers provide significant, complementary information at all stages of reionization, as ${\rm FoM}_{\rm PS2D} < {\rm FoM}_{\rm PS2D+Betti}$. This improvement is most pronounced (around 0.4 dex) at low-to-mid neutral fractions ($\bar{x}_{\rm HI} < 0.6$). Similarly, the middle panel shows the performance of Betti numbers alone. As expected, this ratio is also consistently below zero, indicating that the PS2D adds crucial information not captured by Betti numbers, especially at high neutral fractions ($\bar{x}_{\rm HI} > 0.6$) where the red line dips lowest. Moreover, it is worth noting that for $\bar{x}_{\rm HI} < 0.2$, Betti numbers alone yield similar FoM than PS2D+Betti. The right panel is the most revealing, showing the effect of adding the bispectrum (PS2D+Betti+Bispec) relative to our reference. In agreement with Fig. \ref{fig:FoMdistribs} and Table \ref{tab:FoMresults}, the PS2D+Betti+Bispec only mildly improves on PS2D+Betti. However, this improvement depends on the neutral fraction. At low-to-mid neutral fractions ($\bar{x}_{\rm HI} \lesssim 0.6$), the ratio is centred around zero, indicating that on average the bispectrum adds no relevant information. The result observed for our Late model ($\bar{x}_{\rm HI} \approx 0.70, 0.56, 0.39$) corresponds to one of the dark crosses below 0 in the central \barxHI{} bin. However, at high neutral fractions ($\bar{x}_{\rm HI} \gtrsim 0.6$), the binned median trends above zero. This indicates that for highly neutral scenarios, adding the bispectrum in general brings some new information. Still, dark crosses lying below 0 in the right panel shows that adding the bispectrum can be detrimental for some particular cases. This unlucky negative contribution from higher-order statistics was also found in \cite{Semelin2025combining}, using the pixel density function to improve constraints the astrophysical parameters.

\section{Summary and conclusions}\label{sec:conclusion}

In this work, we have used an implicit inference framework to forecast constraints on the reionization history, parameterised by the average neutral fraction \barxHI{} at three distinct redshifts ($z=8.0, 7.2$ and 6.5). We specifically focused on the added value of higher-order summary statistics—the bispectrum (Bispec) and Betti numbers—when combined with the power spectrum, in the context of upcoming SKA-Low observations. The bispectrum, for which we focused on the equilateral and squeezed-limit configurations, probes the three-point correlations and non-Gaussianity of the field, while Betti numbers capture its topological structure, such as the connectivity of ionised or neutral regions in the 21-cm signal.

Our primary finding is that higher-order statistics contain significant, complementary information to power spectrum. To quantify this robustly—a necessary strategy given that the true reionization history is still loosely constrained—we evaluated the figure of merit (FoM) over our entire test set rather than relying on only a few example mocks. This metric, which is inversely related to the posterior volume, showed that the FoM for constraints from the 2D power spectrum (PS2D) alone was consistently and significantly improved by adding Betti numbers (PS2D+Betti). The bispectrum (PS2D+Bispec) also improved constraints over the PS2D alone. This statistical improvement was also visually confirmed by our analysis of the full posterior constraints for our reference mocks (Fig.~\ref{fig:contours_all}), where the PS2D+Betti combination yielded much tighter posteriors than the PS2D alone, for both 100h and 1000h observation times. This highlights that the topological and three-point information is largely independent of the two-point information in the power spectrum.

However, we also find that the constraining power of a given statistic is not universal, but depends on the reionization state itself. This was most evident with the bispectrum. Our analysis of the full test set (Fig.~\ref{fig:FoMratio}) showed that while PS2D+Bispec improves on PS2D, its benefit varies. The full PS2D+Betti+Bispec combination, which we tested on our reference mocks, provided the tightest constraints for the Early and Fiducial models, but it degraded the constraints for the Late mock (Fig.~\ref{fig:contours_all}). Fig.~\ref{fig:FoMratio} confirmed this trend, showing that the bispectrum adds information in low-to-mid neutral fraction scenarios ($\bar{x}_{\rm HI} \lesssim 0.6$) but can degrade the FoM in highly neutral regimes. This is a critical finding, as it implies that the optimal choice of summary statistics for 21-cm cosmology is state-dependent and must be carefully considered for the specific epoch being probed.

Our analysis, while comprehensive in its comparison of statistics, has limitations. We have not included other significant sources of observational contaminants, such as foreground residuals or data calibration errors. These are known to be important; for example, \citet{greig2024inferring} showed that foreground mitigation could reduce the constraints from the PS2D by a factor of a few. While we refrained from including them here to reduce the complexity of the data model, we acknowledge their importance. With several state-of-the-art foreground mitigation methods being developed \citep[e.g.,][]{mertens2024gprML,acharya2024gprML}, a crucial next step will be to understand the impact of their residuals on non-Gaussian statistics, a task that is not yet well understood and is beyond the scope of this work.

Furthermore, our inference framework was trained and tested using the same simulation code, \textsc{21cmFAST}. It is possible, and indeed likely, that the true universe is not perfectly represented by this simulation. This ``model misspecification'' is a known challenge \citep[e.g.][]{Zhou22}. Classical likelihood-based studies \citep[e.g.,][]{greig201521cmmc,greig2017simultaneously,ghara2020constraining,ghara2021constraining,giri2022imprints,schneider2023cosmological} often account for this by including a ``fudge factor'' or ``modelling error'' term to account for theoretical uncertainties, including differences between different simulation codes; this however delivers conservative posterior contours, which weaken the constraining power of the analysis, and is not guaranteed to produce accurate inferences. Our current SBI framework does not include such a term, and has not been tested on observations from different simulators. Exploring the impact of model misspecification on our higher-order statistics constraints is a critical problem that we will explore in a future paper.

A key motivation for our implicit inference approach is the nature of the reionization history itself. The 21-cm signal is a direct probe of the IGM, and $\bar{x}_{\rm HI}$ is one of the most important properties to infer. However, the reionization history is a derived quantity of the underlying astrophysical model, not a direct input parameter. This makes it challenging to employ classical MCMC methods, which are designed to sample input parameters. Consequently, previous studies \citep[e.g.,][]{ghara2020constraining,ghara2021constraining,ghara2025constraints,greig2021exploring,greig2021interpreting} interpreting the available upper limits on the 21-cm power spectrum have typically inferred the ionising source parameters first, and then mapped those posterior constraints onto a derived reionization history. While some new frameworks have been developed to parameterise IGM properties directly \citep[e.g.,][]{mirocha2022galaxy,ghara2024probing}, they are at an early stage and do not yet capture the complex mechanisms, such as feedback, that modulate the reionization scenarios. Our SBI framework bypasses this step by learning the latent space that maximises the mutual information between the reionization history and the observed data, thereby directly inferring the reionization history and optimising the information extraction.

Our work joins a growing body of research using implicit inference to move beyond the power spectrum. For instance, \citet{Semelin2025combining} also demonstrated the power of combining statistics, showing that SBI could effectively merge the power spectrum and the Pixel Distribution Function (PDF) to achieve tighter constraints than either statistic individually. This parallels our own findings with the PS2D+Betti combination, which similarly combined a Fourier-space and a real image-space summary statistic. Other studies have focused on different non-Gaussian compressors, such as applying SBI to the full 3D image compressed by a 3D Convolutional Neural Network \citep[CNN, in][]{Zhao2022delfi} or, more recently, by a Wavelet Scattering Transform \citep[WST,][]{Zhao2024wavelet}. These works found that non-Gaussian compressors outperformed an SBI analysis applied to the power spectrum alone. These studies, along with our own, converge on a key conclusion: while SBI is a powerful framework, its true potential is unlocked by combining higher-order statistics (like Betti numbers, bispectra, or WST), that capture the complex, topological information of reionization, with the power spectrum.
%while SBI is a powerful framework, its true potential is unlocked by applying it to non-Gaussian statistics (like Betti numbers, bispectra, or WST) that capture the complex, topological information of reionization that the power spectrum misses.

Finally, our implicit inference framework is one of several promising deep learning approaches. In a recent, related study by our collaboration \citep[][]{de2025exploring}, we directly compared different deep learning methodologies, testing various architectures including CNNs, Multi-Layer Perceptron (MLP) Mixers, and an SBI framework. That work focused on the task of inferring $\bar{x}_{\rm HI}$ from the 2D power spectrum alone. We found that the SBI approach was highly competitive, achieving a performance (e.g., $R^2 \approx 97.4\%$) nearly on par with the best-performing, specialized 2D CNNs. This reinforces our finding that implicit inference is a robust and effective method for this inference task. The fact that our collaboration has already demonstrated its performance on the PS2D alone further supports the present results, which show that its predictive power is enhanced when incorporating higher-order summary statistics.

In conclusion, we have demonstrated that an implicit inference framework capable of combining diverse summary statistics is a powerful and necessary tool for 21-cm cosmology. The combination of the power spectrum with non-Gaussian probes—specifically the Betti numbers and, in most regimes, the bispectrum—provides significantly tighter and more robust constraints on the reionization history than the power spectrum alone. This hybrid, multi-statistic approach, which our work validates, will be crucial for maximizing the scientific return from the 21-cm tomographic data expected from the SKA-Low.

\section*{Acknowledgements}

NC acknowledges support by the Swiss National Science Foundation under the Starting grant ``Deep Waves'' (218396). SKG acknowledges the support by Olle Engkvist Stiftelse (grant no. 232-0238).
The authors acknowledge access to Alps at the Swiss National Supercomputing Centre (CSCS), Switzerland, under the SKA share with the project IDs sk014 and sk030. The authors acknowledge funding from the Spark grant CRSK-2\_228671  from the Swiss National Science Foundation, as well as support from the Sweden SKA Regional Center ({\sc sweSRC}) node operated by Onsala Space Observatory in collaboration with Chalmers e-Commons. The Onsala Space Observatory national research infrastructure is funded through the Swedish Research Council grant No 2019-00208. DP is supported by the Swiss National Science Foundation.

%%%%%%%%%%%%%%%%%%%%%%%%%%%%%%%%%%%%%%%%%%%%%%%%%%
\section*{Data Availability}
 
% The datasets created for this article and the trained models will be shared upon request to the authors.
The simulations for the training set were produced with public versions of \textsc{21cmFAST} \citep{murray202021cmfast} and \textsc{Tools21cm} \citep[][]{giri2020tools}. The specific training dataset used in this study is available from the authors upon reasonable request.



\bibliographystyle{mnras}
\bibliography{example} % if your bibtex file is called example.bib

%\begin{thebibliography}{99}
%\end{thebibliography}

%%%%%%%%%%%%%%%%%%%%%%%%%%%%%%%%%%%%%%%%%%%%%%%%%%

%%%%%%%%%%%%%%%%% APPENDICES %%%%%%%%%%%%%%%%%%%%%

\appendix

\section{Posteriors from higher-order statistics alone}
\label{app:hosalone}
\begin{figure}
    \centering
    \includegraphics[width=0.98\linewidth]{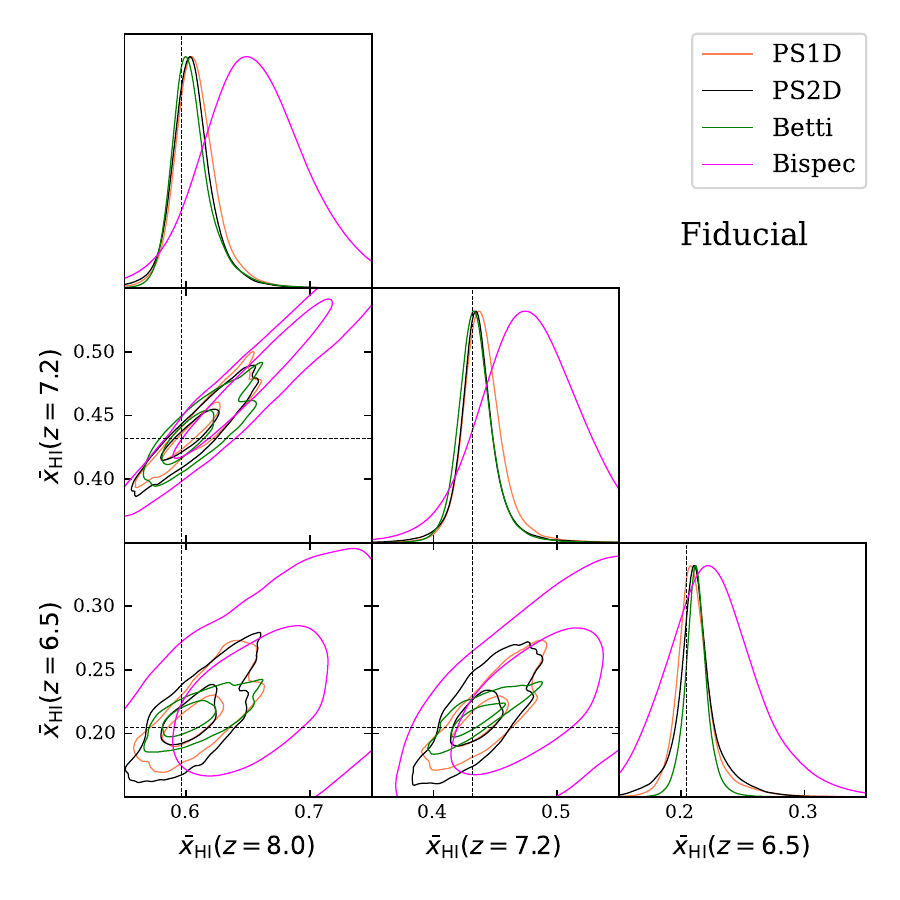}
    \caption{Posterior constraints on the 3D reionization history ($\bar{x}_{\rm HI}(z=8.0)$, $\bar{x}_{\rm HI}(z=7.2)$, $\bar{x}_{\rm HI}(z=6.5)$) for the Fiducial mock at 1000h of noise level. We show the results for each statistic alone: PS1D (coral), PS2D (black), Betti (green) and Bispec (pink).}
    \label{fig:alonehos_posteriors}
\end{figure}

In Section \ref{sec:results_refposterior} we have compared the constraints on reionization for our reference scenarios from different combinations of statistics. For completeness, here we compare the constraints from each statistics alone. In Fig. \ref{fig:alonehos_posteriors} we show the posteriors for PS1D, PS2D, Betti and Bispec on the Fiducial reference mock. We note that the bispectrum is significantly less informative than the others statistics, while the Betti numbers yield slightly tighter constraints than the 2-pt statistics. These observations are in agreement with the global FoM trends shown in Fig. \ref{fig:FoMdistribs}. 

%%%%%%%%%%%%%%%%%%%%%%%%%%%%%%%%%%%%%%%%%%%%%%%%%%

% Don't change these lines
\bsp	% typesetting comment
\label{lastpage}
\end{document}